\documentclass[journal=jctcce,manuscript=article]{achemso}
\setkeys{acs}{articletitle = true}
\usepackage[version=3]{mhchem} 

\usepackage{graphicx}
\usepackage{caption}
\usepackage{natbib}
\usepackage{enumitem}
\usepackage{amsmath}
\newcommand{\angstrom}{\mbox{\normalfont\AA}}

\makeatletter
\def\@email#1#2{%
 \endgroup
 \patchcmd{\titleblock@produce}
  {\frontmatter@RRAPformat}
  {\frontmatter@RRAPformat{\produce@RRAP{*#1\href{mailto:#2}{#2}}}\frontmatter@RRAPformat}
  {}{}
}%
\makeatother

\author{Jaehyeok Jin}
 \email{jj3296@columbia.edu}
\author{David R. Reichman}%
 \email{drr2103@columbia.edu}
\affiliation{Department of Chemistry, Columbia University,
3000 Broadway, New York, New York 10027, USA}%

\title[An \textsf{achemso} demo] {Perturbative Expansion in Reciprocal Space: Bridging Microscopic and Mesoscopic Descriptions of Molecular Interactions}
\begin{document}
\begin{abstract}
Determining the Fourier representation of various molecular interactions is important for constructing density-based field theories from a microscopic point of view, enabling a multiscale bridge between microscopic and mesoscopic descriptions. However, due to the strongly repulsive nature of short-ranged interactions, interparticle interactions cannot be formally defined in Fourier space, which renders coarse-grained approaches in \textit{k}-space somewhat ambiguous. In this paper, we address this issue by designing a perturbative expansion of pair interactions in reciprocal space. Our perturbation theory, starting from reciprocal space, elucidates the microscopic origins underlying zeroth-order (long-range attractions) and divergent repulsive interactions from higher-order contributions. We propose a systematic framework for constructing a faithful Fourier space representation of molecular interactions, capturing key structural correlations in various systems, including simple model systems and molecular coarse-grained models of liquids. Building upon the Ornstein-Zernike equation, our approach can be combined with appropriate closure relations, and to further improve the closure approximations, we develop a bottom-up parametrization strategy for inferring the bridge function from microscopic statistics. By incorporating the bridge function into the Fourier representation, our findings suggest a systematic, bottom-up approach to performing coarse-graining in reciprocal space, leading to the systematic construction of a bottom-up classical field theory of complex aqueous systems. 
\end{abstract}

\begin{center}
\textbf{TOC Graphic} {($3.25'' \times 1.52''$)}\\ 
\end{center} 
\begin{center}
\includegraphics[width=3.25in]{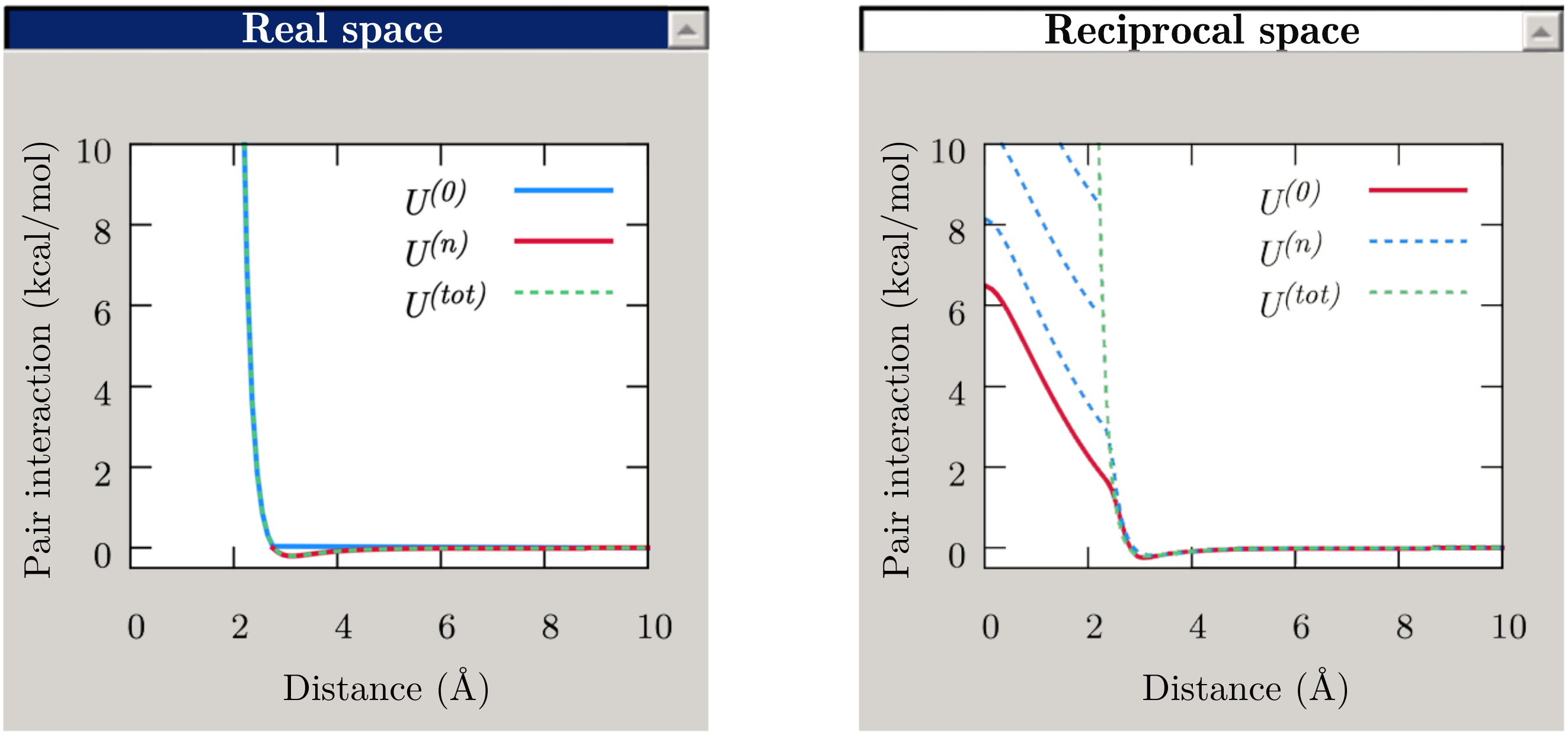}
\end{center}
\clearpage

\section{Introduction} \noindent
In the pursuit of understanding chemical and physical processes in nature, particle-based computer simulations have played a crucial role in exploring microscopic phenomena that are often inaccessible or limited by experimental techniques.\cite{frenkel2001understanding,allen2017computer} Multiscale simulation, in particular, offers a practically appealing approach to efficiently investigate a wide range of molecular processes of interest. Among various multiscale approaches, "bottom-up" multiscale models are constructed from {the statistics of finer-grained systems to study larger length and longer time scale phenomena beyond the reach of finer-grained simulations.}
The primary goal of bottom-up methodologies is to retain the essential microscopic physics in a simplified model, which is crucial for the accuracy and predictability of such multiscale models {across different spatiotemporal scales}. For example, molecular coarse-grained (CG) models can enhance the computational efficiency of fully atomistic simulations while preserving accuracy in terms of structure and dynamics in various settings,\cite{noid2013perspective,saunders2013coarse,jin2022bottom,noid2023perspective} including liquids, materials, and biological systems. However, despite the accuracy and efficiency of these bottom-up models, they are inherently limited by the particle-level description. In other words, the computational cost of performing multiscale simulation scales with the number of particles, and there exists a practical limit where particle-level simulations are no longer efficient enough. 

Given this intrinsic limitation, it is natural to construct a multiscale model by employing a field-theoretical description that goes beyond the particle-level representation.\cite{kardar2007statistical} Rather than a particle-level simulation, the utilization of single or multiple field(s) is expected to significantly extend the spatiotemporal scales of the resultant model by reducing computational overhead. To note, in this direction, Fredrickson and coworkers have demonstrated the efficient sampling of phase diagrams and corresponding equilibrium morphologies of complex soft matter, e.g., polymers.\cite{fredrickson2002field,fredrickson2006equilibrium,fredrickson2023field} In addition, field-theoretic simulations can be used to efficiently recover the free energy {and directly estimate the chemical potential}, which can be highly demanding when approached from a particle-level simulation using enhanced sampling.\cite{bernardi2015enhanced,valsson2016enhancing} Even at the canonical level, field-theoretical approaches enable efficient multiscale simulations. A notable example is the hybrid field-particle simulations for solvated systems,\cite{sides2006hybrid,daoulas2006single,milano2009hybrid,muller2011studying} where numerous solvent particles can be represented as a simple field. However, by construction, these field variables are often derived from simple analytical approximations or \textit{ad hoc} approaches, e.g., phenomenological models such as Edwards' Hamiltonian.\cite{edwards1965statistical,edwards1966theory} Even though the phenomenological nature of classical field theories allows for capturing important physical phenomena of interest, \textit{ad hoc} classical field-theoretical models are prone to overlooking detailed microscopic information underlying the mesoscopic phenomena, including local microscopic structures in equilibrium morphology \cite{zhang2021emergence} or phase transition for polymers\cite{bates2019stability}, as well as local orientational correlations and nanoscale heterogeneity in liquids.\cite{wang2005unique} Therefore, it is of high importance to develop bottom-up classical field models that improve the accuracy and predictability of multiscale models.

In order to develop a bottom-up classical field model, one first needs to define the field variable. While the choice of field variables can be arbitrary depending on the specific system, the density field variable is naturally derived from the particle description in classical phase space.\cite{hansen2013theory} Furthermore, it serves as one of the key field variables for describing the thermodynamic and dynamical properties of a system. In the case of liquids, the density variable is associated with the classical density functional theory in equilibrium\cite{ramakrishnan1979first,oxtoby2002density} and with the dynamical density functional theory for out-of-equilibrium dynamics\cite{marconi1999dynamic,marconi2000dynamic,archer2004dynamical} at the microscopic level. For phase dynamics in complex systems beyond liquids, such as solidification or dislocation, the density variable is closely related to the phase-field crystal approach, where the time evolution of the density field can correctly capture the mesoscopic field dynamics.\cite{elder2007phase,van2009derivation,emmerich2012phase}

Given the depth of information provided by density fields in chemical and physical systems, it is reasonable to select the density field variable for constructing bottom-up field-theoretical models. This bottom-up approach involves reformulating the particle-based Hamiltonian of particle-level systems (molecules) in terms of the density field variable. While the density field itself does not appear in the particle-level description, the density field operator naturally emerges from the interparticle potential energy as a quadratic form in Fourier (reciprocal) space. {This quadratic form enables the formal introduction of auxiliary fields to further decouple the density operator and sample the static configurations by applying the Hubbard-Stratonovich transformation\cite{hubbard1959calculation} in a manner akin to Edwards' approach.\cite{edwards1965statistical,edwards1966theory}} Hence, expressing the potential energy in reciprocal space provides a rigorous foundation for developing a microscopically-derived classical field theory. 

However, the Fourier space representation {at the density field level faces} a critical caveat when applied to molecular systems since the interparticle potential energy in Fourier space is often ill-defined due to its short-range repulsive, {or} \textit{hard-core}, nature. While certain soft matter systems,\cite{fredrickson2018field} such as polymers, may not exhibit such strong repulsion, most liquids or even simple analytical interactions like the Lennard-Jones potential manifest a hard-core region due to the non-penetrating nature of particle overlap, as suggested by classical perturbation theory. This hard-core repulsion diverges as the distance between particles approaches zero, rendering the description of molecular interactions using Fourier space representation infeasible{, and no corresponding field-theoretical simulation can be implemented in practice.} To {mitigate} this numerical issue, several \textit{ad hoc} approaches have been developed that truncate the hard-core repulsion, for example the Lennard-Jones potential, at short distances either manually\cite{singer1958use,laird1991quantum} or using methods such as Gaussian process regression.\cite{bartok2010gaussian,john2017many,deringer2021gaussian} However, given the diverse and complex nature of molecular interactions, these \textit{ad hoc} treatments may not be generally applicable as they might fail to accurately capture the wide range of potential interaction profiles. {This challenge becomes more pronounced when investigating large-scale collective phenomena beyond the molecular level using CG descriptions, as the underlying effective interaction at the CG level is often unknown.} Therefore, it is essential to establish a comprehensive and universal approach to construct a faithful Fourier representation of molecular interactions, independent of specific model parameters. {This approach serves as a general bottom-up CG methodology.}

In this paper, we propose a systematic approach to constructing a Fourier space representation of molecular interactions, which constitutes a fundamental step toward deriving a bottom-up field-theoretical model. Since a divergent hard-core repulsion results in an ill-posed problem, our approach is to derive a systematic perturbative expansion of this hard-core repulsion up to finite orders. In turn, our central objective shares similarities with classical perturbation theory, albeit at different levels of description. Classical perturbation theory in real space considers the hard sphere-like repulsion as the zeroth-order reference interaction, while treating the remaining long-range attractive interactions as perturbations to this reference interaction.\cite{zwanzig1954high,barker1967perturbation,weeks1971role} Motivated by this notion, we aim to extend a similar perturbation idea to reciprocal space. In this framework, the long-range attractive interactions might serve as the reference in reciprocal space, while the hard sphere-like repulsion could be treated as a series of perturbations. 

To the best of our knowledge, such a systematic methodology to address complex and divergent interactions in reciprocal space in the context of field-theoretic coarse-graining remains absent in the existing literature. Nevertheless, readers familiar with the history of liquid-state physics may notice that our approach bears a resemblance to the mode expansion approach of Chandler and Andersen.\cite{andersen1970mode,chandler1971mode,andersen1971mode} While the similarity lies in constructing a perturbative expansion in reciprocal wavevector space, significant distinctions are worth highlighting. Firstly, the core emphasis of mode expansion is to evaluate thermodynamic properties (such as the free energy\cite{andersen1970mode} or activity coefficients\cite{chandler1971mode}) of ionic solutions, assuming a well-defined Fourier component of the potential energy. Conversely, our approach aims to establish a faithful Fourier transformation for inherently divergent interactions, making it suitable for classical field theory. Another pivotal difference pertains to the formulation of perturbative terms as an infinite series, which will be extensively explored in {Sec. \ref{sec:theory}}. In the remainder of this paper, we will further demonstrate that the essential characteristics of divergent molecular interactions can be accurately captured in reciprocal space.
\section{Theory} \label{sec:theory}
\subsection{Fourier Space Representation of Classical Interacting Particles}
We start by considering homogeneous isotropic liquids. Conventionally, the potential energy of the system with a configuration $\mathbf{R}^N$ can be effectively approximated as the sum of pair potentials between particles $I$ and $J$ separated by a distance of $\mathbf{R}_I-\mathbf{R}_J$:
\begin{equation}
    U(\mathbf{R}^{{N}}) = \frac{1}{2} \sum_{I\neq J} U(\mathbf{R}_I-\mathbf{R}_J),
    \label{eq:1}
\end{equation}
where the pair potentials exhibit spherical symmetry. For the sake of notational convenience to account for all pairs, $\sum_{I \neq J}$ in Eq. (\ref{eq:1}) is rewritten from here on as 
\begin{equation}
    U(\mathbf{R}^{{N}}) = \frac{1}{2} \sum_{I, J} U(\mathbf{R}_I-\mathbf{R}_J)-\frac{N}{2}U(R=0),
    \label{eq:1'}
\end{equation}
where $U(R=0)$ denotes the zero distance energy to cancel out $N$ counts of $I=J$ pairs in $\sum_{I,J}$ in Eq. (\ref{eq:1'}). While Eq. ({\ref{eq:1'}) is formally exact, the self-interaction assumes that the potential energy is well-defined at $R=0$. We will validate and examine this assumption in a later section. Hereafter, we note this self-energy correction term as $U_\mathrm{self}$. In order to represent Eq. (\ref{eq:1'}) in Fourier space, we introduce a density field operator that naturally emerges from the classical phase space density: $\rho(\mathbf{R})=\sum_I \delta(\mathbf{R}-\mathbf{R}_I)$. It is straightforward to show that the Fourier component of $\rho(\mathbf{R})$ becomes 
\begin{equation}
      \tilde{\rho}(\mathbf{k}) = \int d\mathbf{R} \rho(\mathbf{R}) \exp(-i\mathbf{k}\cdot \mathbf{R}) = \sum_I e^{-i\mathbf{k}\cdot\mathbf{R}_I}.
\end{equation}
Since the pair interaction encodes information about two particles (as a pair), it naturally implies a harmonic form in Fourier space
\begin{equation}
    U(\mathbf{R}^{{N}}) = \frac{1}{2}\int d\mathbf{R} d\mathbf{R}' \rho(\mathbf{R})U(\mathbf{R}-\mathbf{R}')\rho(\mathbf{R}') - U_\mathrm{self} =\frac{1}{2} \sum_\mathbf{k} \tilde{\rho}^\ast(\mathbf{k}) \tilde{U}(\mathbf{k})\tilde{\rho}(\mathbf{k}) - U_\mathrm{self},
    \label{eq:3}
\end{equation}
assuming the existence of a mathematically well-defined Fourier transformation of the pair interaction, denoted as $\tilde{U}(\mathbf{k}) := \mathcal{F}[U(\mathbf{R})] = \int_0^\infty U(\mathbf{R}) e^{-i\mathbf{k}\cdot\mathbf{R}} d\mathbf{R}$. 

While Eq. (\ref{eq:3}) may seem physically reasonable, its practical implementation is strictly prohibited for molecular systems. 
Since the presence of particle overlap at short distances leads to a highly repulsive potential that scales larger than $R^{-2}$, the Fourier transformation $ \mathcal{F}[U(\mathbf{R})] = \int_0^\infty U(\mathbf{R}) e^{-i\mathbf{k}\cdot\mathbf{R}} d\mathbf{R}$ becomes mathematically ill-defined, posing significant numerical challenges. In this regard, Fourier representations of classical interactions have been applied to only a handful of model interactions with weak repulsion profiles, e.g., the Gaussian Core Model (GCM)\cite{stillinger1976phase} or the Yukawa Model.\cite{yukawa1935interaction} Even though these model systems offer valuable insights about microscopic phenomena, such as polymer solutions\cite{louis2000can} or glass-forming liquids,\cite{ikeda2011glass} it is crucial to resolve the numerical challenges in order to enable realistic representations of more complex molecules. 

\subsection{Ornstein-Zernike Equation in Fourier Space}
Conventional molecular mechanics approaches are typically formulated in real space. However, for fluids or colloidal systems, the Ornstein-Zernike (OZ) equation\cite{ornstein1914influence} provides a clear Fourier space representation of various correlation functions. We are particularly interested in the OZ equation for a uniform, homogeneous system, which can be expressed in the real space representation as follows:
\begin{equation}
    h(\mathbf{R})=c(\mathbf{R})+\rho \int c(|\mathbf{R}-\mathbf{R}'|)h(\mathbf{R}')d\mathbf{R}'.
    \label{eq:4}
\end{equation}
In Eq. (\ref{eq:4}), $h(\textbf{R})$ represents the total correlation function, which is defined as a difference between the radial distribution function (RDF), $g(\textbf{R})$, and its random value (unity) via $h(\textbf{R})=g(\textbf{R})-1$. The function $c(\textbf{R})$ corresponds to the direct correlation function. In short, the OZ equation describes how both direct and indirect correlations in liquid can influence the structural correlation. In the case of homogeneous systems, the effect of indirect correlation becomes a convolution integral, and this provides a convenient starting point to introduce the Fourier transform in reciprocal space through an algebraic expression:
\begin{equation}
    \tilde{h}(\mathbf{k}) = \frac{\tilde{c}(\mathbf{k})}{1-\rho \tilde{c}(\mathbf{k})}.
    \label{eq:OZ-k}
\end{equation}
Alternatively, Eq. (\ref{eq:OZ-k}}) can be expressed using the structure factor, which is defined as
\begin{equation}
    S(\textbf{k}) = 1+ \rho \int d\mathbf{r} e^{-i\mathbf{k}\cdot \mathbf{R}} [g(\mathbf{R})-1]
\end{equation}
and is equivalent to $S(\textbf{k}) = 1+\rho \tilde{h}(\textbf{k})$. The OZ equation in Eq. (\ref{eq:OZ-k}) then connects $\tilde{c}(\textbf{k})$ and $S(\textbf{k})$ in reciprocal space through a relatively simpler form:
\begin{equation}
    S(\textbf{k}) = \frac{1}{1-\rho \tilde{c}(\textbf{k})}.
    \label{eq:Oz-k2}
\end{equation}
However, due to the infinite hierarchy of correlations present in the OZ equation, an approximate closure relation is necessary to truncate the hierarchy. In general, closures are approximate relationships obtained by considering precise diagrammatic expansions of $g(\textbf{R})$ while neglecting certain diagrams.\cite{hansen2013theory} By introducing a bridge function $B(\mathbf{R})$, which includes the sum of all non-nodal elementary diagrams, the closed relationship between $c(\textbf{R})$ and $g(\textbf{R})$ can be expressed as
\begin{equation}
    c(\mathbf{R})=h(\mathbf{R})-\ln g(\mathbf{R})-\beta U(\mathbf{R})+B(\mathbf{R}),
\end{equation}
 with $\beta^{-1}=k_BT$. Determining $B(\textbf{R})$ for a given system requires estimating diagrammatic expansions of the pair functions, which is a highly challenging task and not computationally practical. Henceforth, several approximation schemes have been introduced to simplify the bridge function while capturing the essential microscopic physics of the structural correlation. To note, among the common approximations, the random phase approximation (RPA), or often referred to as the mean spherical approximation, is the simplest closure, given by $c(\mathbf{R})=-\beta U(\mathbf{R})$.\cite{hansen2013theory} A more refined description can be obtained through the hypernetted chain (HNC) approximation, where $B(\mathbf{R})=0$.\cite{morita1958theory} As our main focus is on liquids, we will primarily utilize the RPA and HNC closures. For example, we do not explicitly consider the Percus-Yevick closure in this work because it is mainly applicable to short-range hard interactions.\cite{percus1958analysis} Nevertheless, we will provide a more general discussion of the microscopically estimated bridge function in {Sec. \ref{sec:bridge}}.

\subsection{Perturbation Theory in Reciprocal Space}
As introduced, the HNC approximation for spherical systems results in the following form of the pair potential 
\begin{equation}
    U(R)=\frac{1}{\beta} \left( h(R)-\ln g(R)-c(R)\right).
    \label{eq:HNC}
\end{equation}
Equation (\ref{eq:HNC}) provides further insight into the numerical issues associated with the Fourier representation of divergent repulsive interactions. Due to the hard-core repulsion, as $R$ approaches 0, $g(R)\rightarrow 0$, implying $\lim_{R\rightarrow 0} U(R) \rightarrow \infty$. Therefore, in order to obtain a well-defined Fourier transformation of $U(R)$, we aim to understand the divergent behavior of $\ln g(R)$ at short distances by expressing it in a perturbative form using a Taylor series. Since $g(R)=1+h(R)$, we can express $\ln g(R)$ as
\begin{equation}
    \ln g(R)=\sum_{n=1}^\infty (-1)^{n+1} \frac{[h(R)]^n}{n}.
    \label{eq:Taylor}
\end{equation}
Note that the first term $n=1$ from Eq. (\ref{eq:Taylor}) corresponds to the RPA description, where the direct correlation in real space can be written as
\begin{equation}
    c(R)\approx h(R)-h(R)-\beta U(R)+\mathcal{O}[h(R)^2].
    \label{eq:RPA1}
\end{equation}
Since Equation (\ref{eq:RPA1}) has a well-defined Fourier component $\tilde{U}^{(0)}$, then Eq. (\ref{eq:Oz-k2}) can be used to find $\tilde{U}^{(0)}=\frac{1}{\rho \beta} \left(\frac{1}{S(k)-1}\right)$ in reciprocal space. In other words, the RPA is the zeroth-order truncation of the HNC closure that gives a well-defined Fourier component. Nevertheless, the RPA does not provide an accurate description of structural correlations at short wavelengths due to the missing correlations $\mathcal{O}[h(R)^2]$. In this sense, a higher-order perturbative expansion gives the following correction terms: 
\begin{equation}
    U(R) = -\frac{c(R)}{\beta} +\frac{1}{\beta}\left( \frac{h^2(R)}{2} - \frac{h^3(R)}{3}+\cdots \right)
    \label{eq:R-Taylor}
\end{equation}
in real space, and consequently,
\begin{equation}
    \tilde{U}(k)=\frac{1}{\rho \beta}\left(\frac{1}{S(k)}-1\right)+\frac{1}{\beta}\left(\frac{\tilde{h}^2(k)}{2}-\frac{\tilde{h}^3(k)}{3} +\cdots \right)
    \label{eq:k-Taylor}
\end{equation}
in reciprocal space. For clarity, hereafter we denote the $\textit{n}$-th correction term in Eq. (\ref{eq:k-Taylor}) as  $\tilde{U}^{(n)}(k) := \frac{(-1)^n}{\beta}\frac{\tilde{h}^n(k)}{n}$, i.e., 
\begin{equation}
    \tilde{U}(k)=\tilde{U}^{(0)}(k)+\sum_{n=2}^{\infty}\tilde{U}^{(n)}(k).
    \label{eq:Perturb}
\end{equation}

{We note that Eq. (\ref{eq:Taylor}) converges only when $|h(R)|<1$, corresponding to low-density systems. Hence, for high-density systems where $g(R)>2$, we provide a generalization of Eqs. (\ref{eq:Taylor})--(\ref{eq:k-Taylor}) in Appendix A. To validate the effectiveness of our approach, we will apply our theory to molecular systems under both low- and high-density conditions in Sec. \ref{sec:three}.}

\subsection{Link to Classical Perturbation Theory}

In the previous section, we demonstrated that under the HNC closure, the pair interaction can be expressed as a type of perturbative expansion of the RPA closure. However, we would like to note that this perturbative expansion is tractable in reciprocal space and differs from classical perturbation theory. In this section, we analyze the perturbative structure of Eq. (\ref{eq:Perturb}) in order to understand the physical origins underlying these perturbation terms.

\subsubsection{Perturbative Structure}

Classical perturbation theory is originally designed to calculate the free energy and other thermodynamic properties in a perturbative manner. Examples of such approaches include the Barker-Henderson \cite{barker1967perturbation} and the Weeks-Chandler-Andersen\cite{weeks1971role} perturbation theories. In these theories, the pair interaction is typically split into two components in real space, i.e., $U(R)=U^{(0)}(R)+U^{(1)}(R)$. Then, the canonical partition function
\begin{equation}
    \mathcal{Z}=\int\cdots\int d \mathbf{R}_1\cdots d \mathbf{R}_n \exp[-\beta(U^{(0)}(R)+U^{(1)}(R))],
    \label{eq:CP1}
\end{equation}
can be further factorized into
\begin{equation}
    \mathcal{Z}=\mathcal{Z}^{(0)}\langle \exp[-\beta U^{(1)}(R)]\rangle,
    \label{eq:CP2}
\end{equation}
where $\mathcal{Z}^{(0)}$ is the partition function derived from $U^{(0)}$ as $\mathcal{Z}^{(0)}=\int\cdots\int d \mathbf{R}_1\cdots d \mathbf{R}_n \exp[-\beta U^{(0)}(R)]$. The multiplicative factorization of $\mathcal{Z}$ implies that the configurational contribution to the Helmholtz free energy, $A=-\frac{1}{\beta}\ln \mathcal{Z}$, can be decomposed as
\begin{equation}
    \beta A = \beta A^{(0)}+\beta A^{(1)},
\end{equation}
which retains a perturbative structure in the free energy.

In our case, the total potential energy in $\mathbf{R}^{{N}}$ configurational space can be represented in the Fourier space as
\begin{equation}
    U(\mathbf{R}^{{N}})=\frac{1}{2}\sum_k \tilde{\rho}^*(k)\tilde{U}(k)\tilde{\rho}(k),
\end{equation}
where $U_\mathrm{self}$ is omitted here for simplicity. Since Eq. (\ref{eq:Perturb}) is linearly decomposed in $k$-space, i.e., $\tilde{U}(k)=\tilde{U}^{(0)}(k)+\sum_n\tilde{U}^{(n)}(k)$, the canonical partition function can also be factorized similarly into Eqs. (\ref{eq:CP1}) and (\ref{eq:CP2}) due to 
\begin{equation}
    \mathcal{Z}= \int\cdots\int d \mathbf{R}_1\cdots d \mathbf{R}_{{N}}\exp\left[-\frac{\beta}{2}\sum_k \left\{ \tilde{\rho}^* (k)\left( \tilde{U}^{(0)}(k)+\sum_n\tilde{U}^{(n)}(k) \right) \tilde{\rho}(k)\right\} \right],
    \label{eq:perturbation}
\end{equation}
{where $\tilde{\rho}(k)$ is the Fourier expression of the microscopic density operator, i.e., $\rho(\textbf{R})=\sum_I \delta(\mathbf{R}-\mathbf{R}_I)$, and hence $\tilde{\rho}(k)$ is a function of both the $k$ vector and CG configurations $\{\mathbf{R}_I\}$: $\tilde{\rho}(k)=\sum_I \exp (-ik\mathbf{R}_I)=\tilde{\rho}(k, \{\mathbf{R}_I\})$. In turn, Eq. (\ref{eq:perturbation}) gives $A = A^{(0)}+\sum_n A^{(n)}$ in reciprocal space.} Therefore, our theory also exhibits a perturbative structure in terms of the free energy, similar to conventional perturbation approaches, and Eq. (\ref{eq:perturbation}) indicates that our perturbation acts on the reciprocal space in the opposite manner. 

Nevertheless, in order to correctly understand the perturbative structure of $\tilde{U}(k)$ based on our decomposition scheme, it is necessary to investigate the physical origins underlying the real space representation of  $\mathcal{F}^{-1}[\tilde{U}^{(0)}(k)]$ and $\mathcal{F}^{-1}[\tilde{U}^{(n)}(k)]$. The first step in this direction would be to examine the relative magnitude of the perturbative terms with respect to the zeroth-order value. For the sake of simplicity, we consider when $R=0$ (real space) and $k=0$ (reciprocal space). In real space, the convergence of $\mathcal{F}^{-1}[\tilde{U}(k)]$ depends on the behavior of $g(R=0)$. If $g(R=0)>0$, then the inverse Fourier transform will converge. This condition is satisfied when $U(R=0)$ is finite, e.g., Yukawa or GCM interactions. In such cases, one could directly estimate the Fourier component without encountering any numerical issues. However, for complex soft-core interactions, the zero-distance value becomes 
\begin{equation}
    \mathcal{F}^{-1}[\tilde{U}(k)]=-\frac{1}{\beta}c(R=0)+\frac{1}{\beta}\sum_n \frac{1}{n}\rightarrow\infty.
    \label{eq:convergence}
\end{equation}

The reciprocal space representation $\tilde{U}(k)$ is slightly more complicated than its real space representation, but if $U(R=0)$ diverges, then $\tilde{U}(k=0)$ should diverge as well. Given the divergent nature of $\tilde{U}(k)$ for hard-core pair interactions, this analysis provides a detailed insight into approximating a faithful Fourier representation of complex pair interactions.

\subsubsection{Physical Origins and Implications}

From Eq. (\ref{eq:Perturb}), the zeroth-order perturbation term corresponds to the RPA closure, which has a well-defined and finite value for any chemical system, as it can be written as $\frac{1}{\rho \beta}\left(1/S(k) -1\right)$. Also, it should be noted that, unlike the zeroth-order term in conventional perturbation theory, $\mathcal{F}^{-1}[\tilde{U}^{(0)}]$ contains both the short- and long-range interactions from $c(R)$.

Therefore, the divergence of $\tilde{U}(k)$ can be attributed to the contribution from higher-order terms. By definition, these higher-order terms are related to the inverse Fourier transformation of the total correlation function, $\tilde{h}(k)$, 
\begin{equation}
    \mathcal{F}^{-1}[\tilde{U}^{(n)}(k)]=\mathcal{F}^{-1}\left[\frac{(-1)^n}{\beta}\frac{\tilde{h}^n(k)}{n}\right],
\end{equation}
indicating that $\tilde{h}(k)$ is central to understanding the behavior of $\mathcal{F}^{-1}[\tilde{U}^{(n)}(k)]$ in reciprocal space. H{\o}ye and Stell previously expressed that $\tilde{h}(k)$ for molecular liquids can be described as a summation of the spherical Bessel functions with prefactors. \cite{sullivan1981evaluation,ho1976dielectric,ho1977kerr} Therefore, in the single-site description, it is reasonable to approximate $\tilde{U}^{(n)}$ for higher-order ($n\gg1$) as the first-order Bessel function while all other peaks are relatively smoothed. In Section 3, we will assess the validity of this approximation for various interaction profiles. Since the inverse Fourier transformation of the Bessel function is a rectangular function, $\mathcal{F}^{-1}[\tilde{U}(k)]$ for higher-order $n\gg1$ results in a square-like well interaction within the first coordination shell in real space.

\begin{center}
\includegraphics[scale=0.12]{"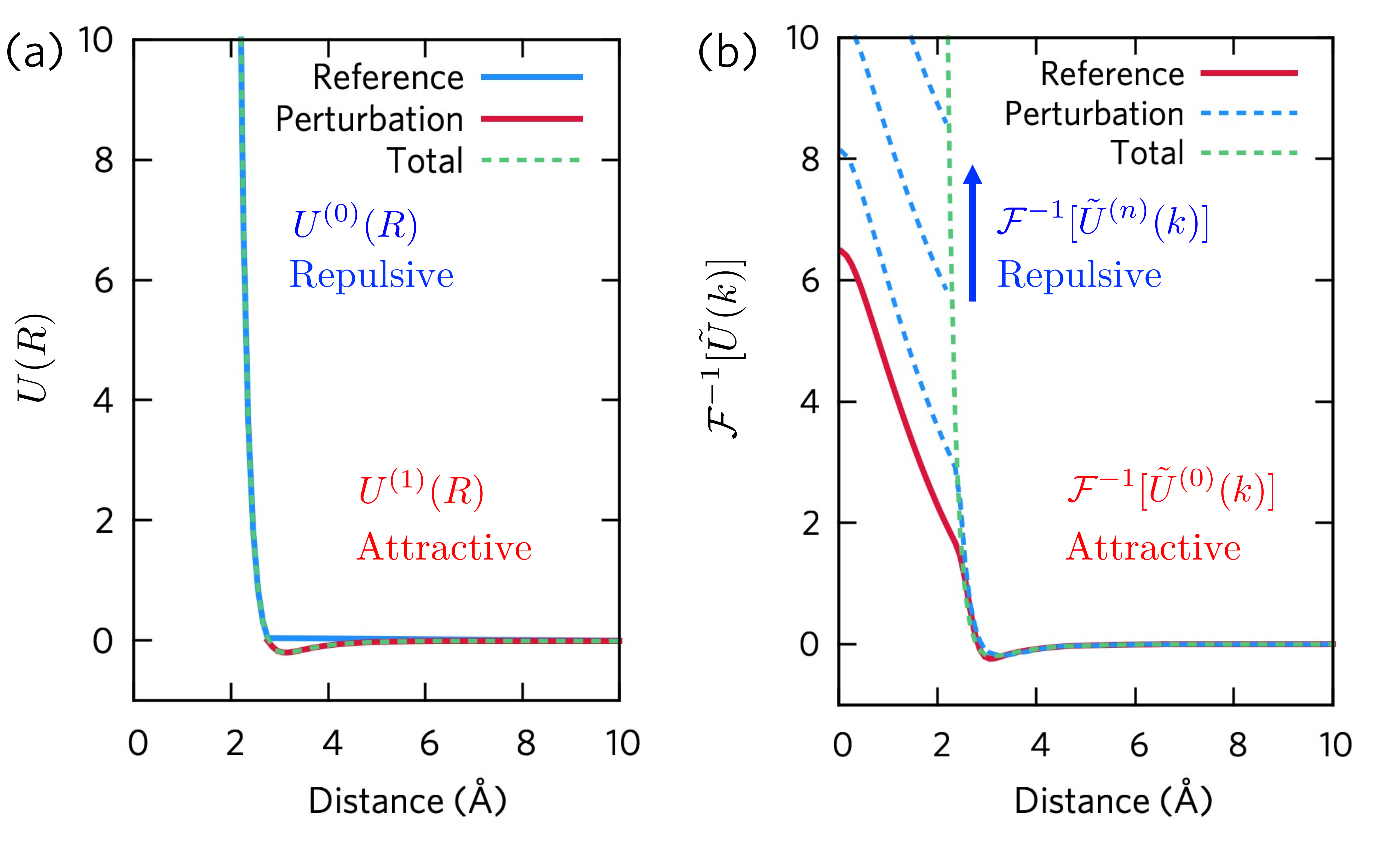"}\\
\textbf{Figure 1}: Schematic illustration of the perturbation expansion of interactions (a) in real space from classical perturbation theory and (b) in reciprocal space proposed in this work
\label{fig:Schematic}
\end{center}

In summary, starting from the perturbation expansion of interactions in reciprocal space, i.e.,  $\tilde{U}(k)=\tilde{U}^{(0)}(k)+\sum_{n=2}^{\infty}\tilde{U}^{(n)}(k)$, the real space representation (i.e., the inverse Fourier transform) of this perturbation implies the following: First, the zeroth-order perturbation $\mathcal{F}^{-1}[\tilde{U}^{(0)}(k)]$ gives the long-range interaction as the reference. Then, the summation of perturbation terms $\mathcal{F}^{-1}[\tilde{U}^{(1)}(k)]$ results in an infinite repulsion at the hard-core region. As discussed in the Introduction, this behavior is completely opposite to classical perturbation theory in real space. In the classical case, the perturbation scheme $U(R)=U^{(0)}(R)+U^{(1)}(R)$ partitions the hard sphere-like repulsion as the reference potential $U^{(0)}(R)$ and the long-range (attractive) interaction $U^{(1)}(R)$. The opposite correspondence between $U^{(1)}(R)$ and $\mathcal{F}^{-1}[\tilde{U}^{(n)}(k)]$ is consistent with the relationship between real space and reciprocal space. Figure 1 summarizes this section. 

\subsection{Implementation {and Parametrization}}
In order to validate our theory, we initially start with a zeroth-order approximation and examine the performance of $\tilde{U}^{(0)}(k)$. The zeroth-order term can be readily obtained through the RPA closure from the structure factor of the target system
\begin{equation}
    \tilde{U}^{(0)}(k)=\frac{1}{\rho \beta}\left(\frac{1}{S(k)}-1\right).
    \label{eq:zeroth}
\end{equation}
Then, we will incorporate the higher-order perturbative term $\tilde{U}^{(n)}(k)$, which can be computed using
\begin{equation}
    \tilde{U}^{(n)}(k) := \frac{(-1)^n}{\beta}\frac{\tilde{h}^n(k)}{n}.
    \label{eq:nth}
\end{equation}

To validate the accuracy of our estimated Fourier space interaction in capturing the overall potential energy, we choose to numerically evaluate $U^{(n)}(R) = \mathcal{F}^{-1}[\tilde U^{(n)}(k)]$ through an inverse Fourier transformation of the estimated $\tilde U^{(n)}(k)$. Subsequently, in order to demonstrate the fidelity of the resultant Fourier representation in capturing the microscopic correlations, we conduct particle-based MD simulations using $U^{(n)}(R)$ and compute the structural correlations. This indirect approach allows us to examine the performance of the Fourier representation. 

\subsubsection{{Bottom-Up Approach: Molecular Systems}}
To demonstrate the versatility of our {bottom-up} approach, we consider three distinct categories of systems due to the diverse nature of realistic hard-core interactions. Firstly, we examine the Gaussian model (GCM), which exhibits a well-defined Fourier representation that can be analytically expressed. By quantitatively assessing our approach using this model, we establish its validity before tackling more challenging systems with hard-core repulsions. Once we have established the validity of our approach, we proceed to apply it to the widely studied Lennard-Jones (LJ) interaction, a well-known model system exhibiting strongly repulsive interactions. By successfully applying our method to the LJ interaction, we demonstrate its potential for extension to complex, realistic systems beyond analytical model interactions.

Building upon the findings from model interactions, we then extend our approach to encompass complex and realistic systems. Considering the utility of field-level interactions in Fourier space for multiscale simulation, the {CG} interactions of molecules would be the ideal target. To this end, we first construct molecular {CG} models from atomistic (microscopic) interactions, where the {CG} superatoms are mapped as a center-of-mass of molecules. The governing conservative interactions between these {CG} sites are {derived from the center-of-mass-level correlations, namely $S(k)$ and $\tilde{h}(k)$, which are obtained from the mapped atomistic trajectory. Utilizing the Fourier space representation for CG molecular liquids, we conducted CG MD simulations at the particle-level by employing the inverse Fourier-transformed real space interactions. Due to the renormalized degrees of freedom during the coarse-graining process, the thermodynamic properties of CG models (e.g., entropy\cite{jin2019understanding} or pressure\cite{dannenhoffer2019compatible}) often deviate from their atomistic references. As a consequence of this challenge, known as the representability issue in CG modeling,\cite{johnson2007representability,wagner2016on,dunn2016van} we have opted to assess the fidelity of perturbative models by estimating structural correlations.}

\subsubsection{{Implication for a Top-Down Approach}} \label{subsec:topdown}
{While our theory was originally designed for constructing a bottom-up density field representation of molecular systems, we would like to emphasize that our method can be seamlessly applied in the opposite direction, commonly referred to as a \textit{top-down} approach. In a top-down approach, one can directly derive the zeroth and higher-order perturbative terms from experimental results. For example, experimental data such as the structure factor $S_\mathrm{exp}(k)$ obtained from X-ray or neutron scattering experiments can serve as a starting point.}

{By inverting the Ornstein-Zernike equation [Eq. (\ref{eq:Oz-k2})], one can directly obtain $\tilde{c}(k)$ from $S_\mathrm{exp}(k)$. Additionally, experimental pair correlations can be determined from $S_\mathrm{exp}(k)$ through inverse Fourier transformation:
\begin{equation}
    g(\mathbf{R}) = 1+ \frac{1}{(2\pi)^3} \int d\mathbf{k} \frac{S(\mathbf{k})-1}{\rho} e^{-i \mathbf{k} \cdot \mathbf{R}}.
\end{equation}
With both $c(R)$ and $h(R)$ at hand, one can utilize Eqs. (\ref{eq:zeroth}) and (\ref{eq:nth}) to construct $\tilde{U}^{(0)} (k)$ and $\tilde{U}^{(n)} (k)$, based solely on experimental observables. While this approach can in principle be applied to arbitrary molecular systems, we leave a detailed top-down implementation for future work as the primary focus of this paper lies in the construction of a microscopic Fourier space representation through a bottom-up approach.
}
\section{Results and Discussions} \label{sec:three}
\subsection{Test System: Gaussian Core Model}

\subsubsection{Setup and Computational Details} 

The GCM \cite{stillinger1976phase} can be described with the interaction form
\begin{equation}
    U_{GCM}(R)=\varepsilon \exp \left(-\frac{R^2}{R_0^2} \right),
    \label{eq:GCM}
\end{equation}
where $\epsilon$ determines the energy scale and $R_0$ represents the length scale of the system. Since Eq. (\ref{eq:GCM}) follows a Gaussian form, the Fourier transform of $U_{GCM}$ can be analytically expressed as a rescaled Gaussian form of 
\begin{equation}
    \tilde{U}_{GCM}(k)=\pi^{\frac{3}{2}}R_0^{3}\varepsilon\exp\left(-\frac{k^2R_0^2}{4}\right)
\end{equation}
in a \textit{k}-space representation. Since the GCM has been shown to accurately reproduce the polymer solution regime, when $\varepsilon^\ast=\varepsilon/k_BT \approx 2$ under the semi-dilute number density regime $\rho^\ast = \rho R_0^3 \ge \frac{3}{4\pi}$. We selected our parameters to also fall within the polymer-like regime, i.e., $\rho^\ast = 0.5$.

We constructed the GCM system with 8000 particles in a cubic box of 62.556 $\angstrom \times$ 62.556 $\angstrom \times$ 62.556 $\angstrom$. To enforce the reduced number density $\rho^\ast=0.5$, we chose $R_0$ to be 1.755 $\angstrom$ in real units. The interaction strength $\varepsilon$ was set to 1.194 kcal/mol, and the system temperature was maintained at 300 K, resulting in $\varepsilon^\ast = 2$. The initial configuration of the GCM system was randomly generated using the Packmol software.\cite{martinez2009packmol} Subsequently, we performed energy minimization to remove any artificial strain on the system. From the energy-minimized configuration, we performed constant $NVT$ dynamics at 300 K using the N\'{o}se-Hoover thermostat\cite{nose1984unified,hoover1985canonical} for 2.5 ns. The last half of the constant $NVT$ run was collected and used to compute the structural correlations and to estimate the Fourier representation using our proposed approach. Throughout the manuscript, the Large-scale Atomic/Molecular Massively Parallel Simulator (LAMMPS) MD engine \cite{plimpton1995fast,brown2011implementing,brown2012implementing} was employed to perform particle-based MD simulations.

\subsubsection{Random Phase Approximation: The zeroth-order term}

From the collected MD trajectory, we initially estimated the zeroth-order perturbation term by employing an RPA closure. The RPA representation was readily obtained by computing the structure factor $S(k)$ from atomistic simulation using
\begin{equation}
    S(k)={\frac{1}{N}}\langle \tilde{\rho}(k)\tilde{\rho}(-k)\rangle.
\end{equation}
Figure 2(a) compares the RPA interaction with the reference $U_{GCM}$ in real space. As expected,
the RPA closure accurately captures the long-range decaying behavior, but there is a noticeable deviation in the short-range repulsive region compared to the reference interaction. This difference corresponds to $\sum_n \mathcal{F}^{-1}[\tilde{U}^{(n)}(k)]$. 

A similar trend was also observed in reciprocal space, where we can directly evaluate the full analytical expression for the reference Gaussian interaction, unlike other hard-core potentials. While the difference is less pronounced in reciprocal space [Fig. 2(b)], we note that the RPA exhibits a slight deviation from the reference $\tilde{U}_{GCM}(k)$, especially as $k\rightarrow0$. The difference in $k$-space can be attributed to the fact that differences at short distances in real space are Fourier transformed in reciprocal space, resulting in smearing over the entire $k$-space. 

Finally, the accuracy of the approximated interaction in terms of structural correlations is depicted in Fig. 2(c) by computing the radial distribution function (RDF) from particle-level simulation using $\mathcal{F}^{-1}[\tilde{U}^{(0)}(k)]$. It is somewhat surprising to observe that the RPA closure yields a rough representation of the reference atomistic RDF. This qualitative agreement suggests that the RPA closure can be utilized as a crude approximation of the complex hard-core repulsion that produces a correct long-range description. Nevertheless, in the short-range regime, the RPA description starts to deviate from the reference correlation, especially at $R=0$. 

\begin{center}
\includegraphics[scale=0.12]{"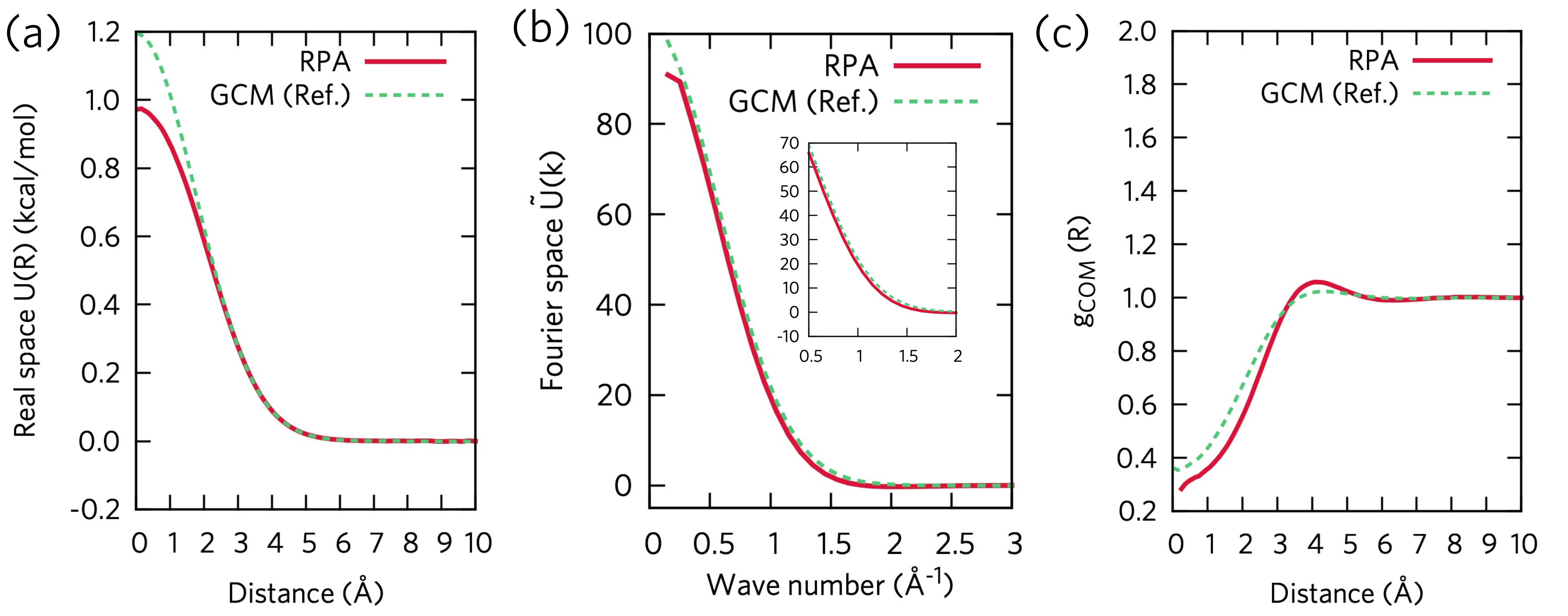"}\\
\textbf{Figure 2}: Performance of RPA (red lines) compared to the reference for the GCM liquid (dashed green lines) in terms of: (a) real space interaction, (b) reciprocal space interaction, and (c) RDF computed from MD simulations in real space.
\label{fig:GCM-RPA}
\end{center}

\subsubsection{Perturbation Theory in Reciprocal Space}

In order to improve the Fourier space representation of the GCM, we consider the $h({R})$ function. By definition, $h({R})$ is derived from the atomistic MD RDF, i.e., $h({R})=g({R})-1$, as illustrated in Fig. 2(c). For each order $n$, we performed a Fourier transform of $h^{n}({R})$ and then estimated $\tilde{U}^{(n)}(k)$ using $(-1)^n\tilde{h}^{n}(k)/(\beta n)$. As can be seen from the finite difference between the reference and RPA interactions in Fig. 2, this perturbation is expected to converge. To further demonstrate this perspective, Fig. 3 illustrates the interaction profile of the perturbative terms [(a)-(c) in real space and (d)-(f) in reciprocal space representations]. Notably, Fig. 3(b) displays the approximated interaction up to the $n$-th perturbative term, and we observe that the overall perturbative interaction converges toward the reference GCM interaction. It is worth noting that this perturbation primarily affects short-range regions, including the zero distance behavior, whereas the long-range interactions are already well captured by the RPA (reference) interaction [Fig. 3(a)]. To clearly illustrate the convergence, Fig. 3(c) compares the interaction value at zero distance, $\mathcal{F}^{-1}[\tilde{U}^{(n)}](R=0)$, for each perturbation. After including up to $n=10$, the interaction value converges to $U_{GCM}(R=0)$. 

A similar agreement can be observed in reciprocal space, as depicted in Fig. 3(d)-(f). At zero wave vector, the $n$-th order perturbative term in reciprocal space converges after $n=10$, accurately capturing the reference $\tilde{U}_{GCM}(k)$, especially at $n=0$. Since both the real and reciprocal space interactions converge after $n\approx10$, we truncated the perturbation at $n=10$ and performed the MD simulation using the truncated perturbative interaction obtained from the inverse Fourier transform. As depicted in Fig. 3(d), the converged interactions in Fourier space yield almost identical structural correlations, indicating that the proposed approach can be applied to hard-core models with strongly repulsive interactions.

\begin{center}
\includegraphics[scale=0.35]{"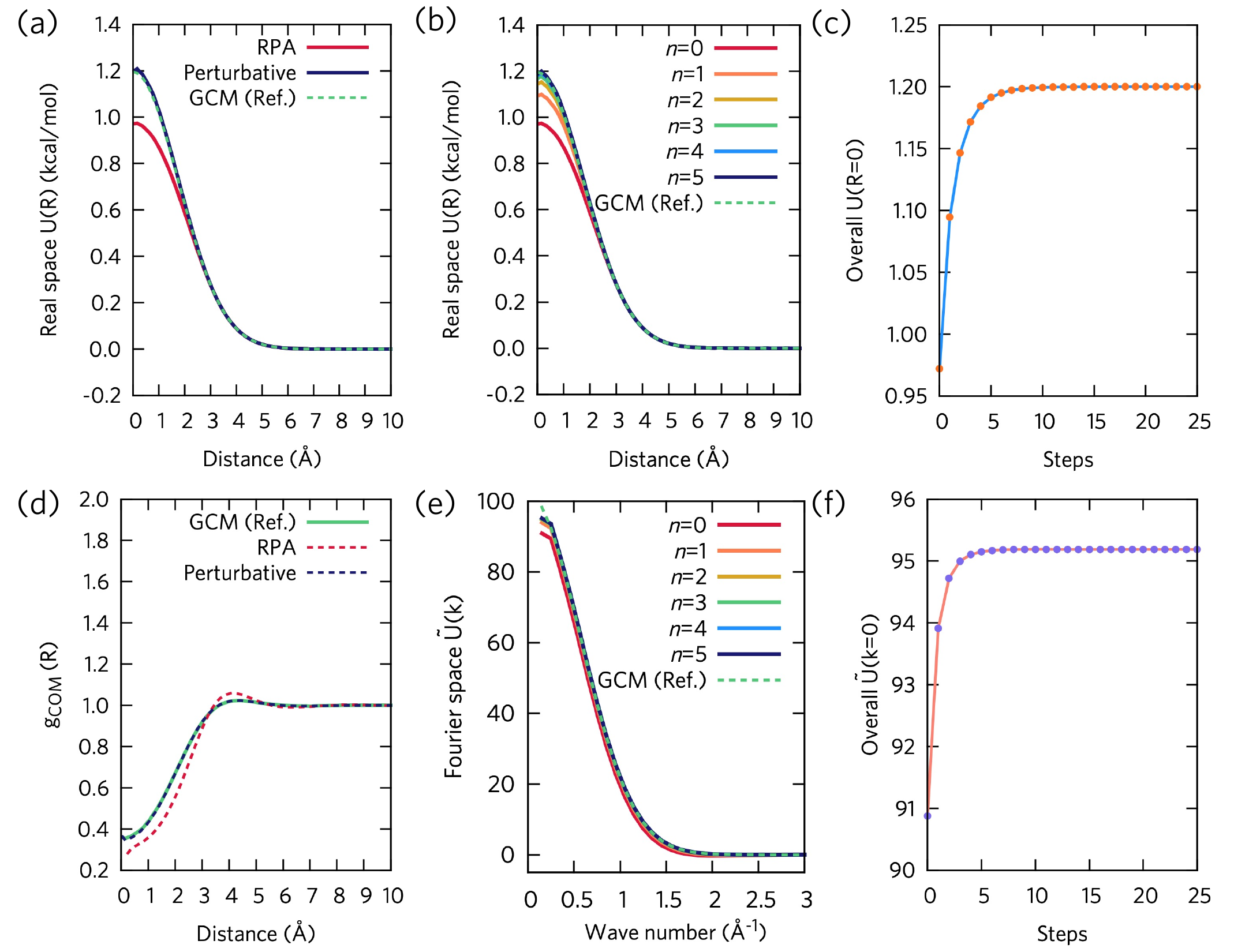"}\\
\textbf{Figure 3}: Perturbative expansion of the analytical GCM liquid. (a) Overview of interactions in real space where the corresponding RDFs are shown in (d). The truncated perturbative interactions up to the $n$-th order are shown for (b) the real space representation and (e) the Fourier space representation. The convergence behavior of each order is depicted in (c) real space at $R=0
$ and (d) Fourier space at $k=0$. 
\end{center}

\subsection{Model System: Lennard-Jones Model} \label{sec:lj}

\subsubsection{Setup and Computational Details} 

After confirming the validity of our approach in the analytical GCM case, we now transition to a simple yet universal hard-core model commonly used to describe molecular systems—the Lennard-Jones LJ model. In contrast to the GCM, the LJ model features a short-range stiff repulsive interaction that scales as $R^{-12}$ from the interaction form $U_{LJ}(R)$,
\begin{equation}
    U_{LJ}(R) = 4\varepsilon \left[ \left( \frac{\sigma}{R}\right)^{12} - \left( \frac{\sigma}{R}\right)^6  \right].
    \label{eq:LJ}
\end{equation}
Hence, a direct Fourier transformation of $U_{LJ}(R)$ is not possible, and the finite perturbation approach may be able to capture the essential correlations arising from the repulsion. In Eq. (\ref{eq:LJ}), the energy scale is denoted by $\varepsilon$, and the length scale is represented by $\sigma$. 

For constructing the effective Fourier representation, since the OZ equation is valid in the liquid regime, we chose supercritical conditions, where the reduced number density is $\rho^\ast=\rho\sigma^3=0.7$, and the reduced temperature is $T^\ast = k_BT/\varepsilon=2.5$ and $ 3.0$. When preparing the reference atomistic simulations, we used the same box size of $62.556 \angstrom$ as in the GCM to ensure consistency and placed 8000 particles, giving $\sigma = 2.777 \angstrom$. The effective energy unit for $\varepsilon$ was chosen as 0.239 and 0.199 kcal/mol at 300 K, corresponding to $T^\ast= 2.5$ and $3$, respectively. The simulation protocol was identical to the GCM case, where a constant \textit{NVT} simulation was performed at the energy-minimized structure for 2.5 ns using the N\'{o}se-Hoover thermostat. 

\subsubsection{Results for $T^\ast=3.0$}

We first constructed the perturbative interaction using the RPA expression to assess if the RPA description could still capture the long-range interaction profile for hard-core systems. Interestingly, Fig. 4(a) clearly demonstrates that even for hard-core systems, the RPA closure can successfully capture the long-range attractions and partially reproduce the short-range repulsion up to a finite value. However, it is important to note that the RPA interaction alone cannot accurately represent the structural correlations of the LJ system, as evident from the RDF computed from MD simulations using only the RPA interaction [Fig. 4(d)]. Notably, while the differences between the RPA and atomistic reference are not pronounced at longer distances (e.g., $R>4 \angstrom$), the overstructured $g(R)$ in the first coordination shell strongly suggests that the short-range treatment of RPA is insufficient for reproducing the reference correlation. Additionally, Fig. 4(d) shows a non-zero value for the pair distribution below the first coordination shell, in contrast to the reference liquid. This penetrating nature is contrary to the expected behavior based on the hard-core nature of the LJ interaction. However, it can be understood by considering the finite repulsion at $U^{(0)}(R)$, which indicates the system experiences a finite interaction at the given temperature. Nevertheless, it is particularly interesting to observe that the RPA model can approximately capture the general shape of the RDF after the second coordination shell for the LJ system, suggesting that the RPA closure can provide an efficient estimate for the long-range structural correlation. 

In order to correct the RPA behavior, we computed the perturbative terms in Fourier space [Fig. 4(f)] and then performed an inverse Fourier transformation in real space [Fig. 4(c)]. At lower order perturbations [e.g., $n=100$ in Fig. 4(c) and (f)], the interaction profile slightly deviates from the Bessel function. However, at higher orders, this peculiarity appears to be averaged out, and it becomes evident that $\tilde{U}^{(n)}$ follows a Bessel function of the first order with the corresponding $U^{(n)}(R)$ exhibiting a square-like repulsion at short distances, as hypothesized earlier. 

\begin{center}
\includegraphics[scale=0.4]{"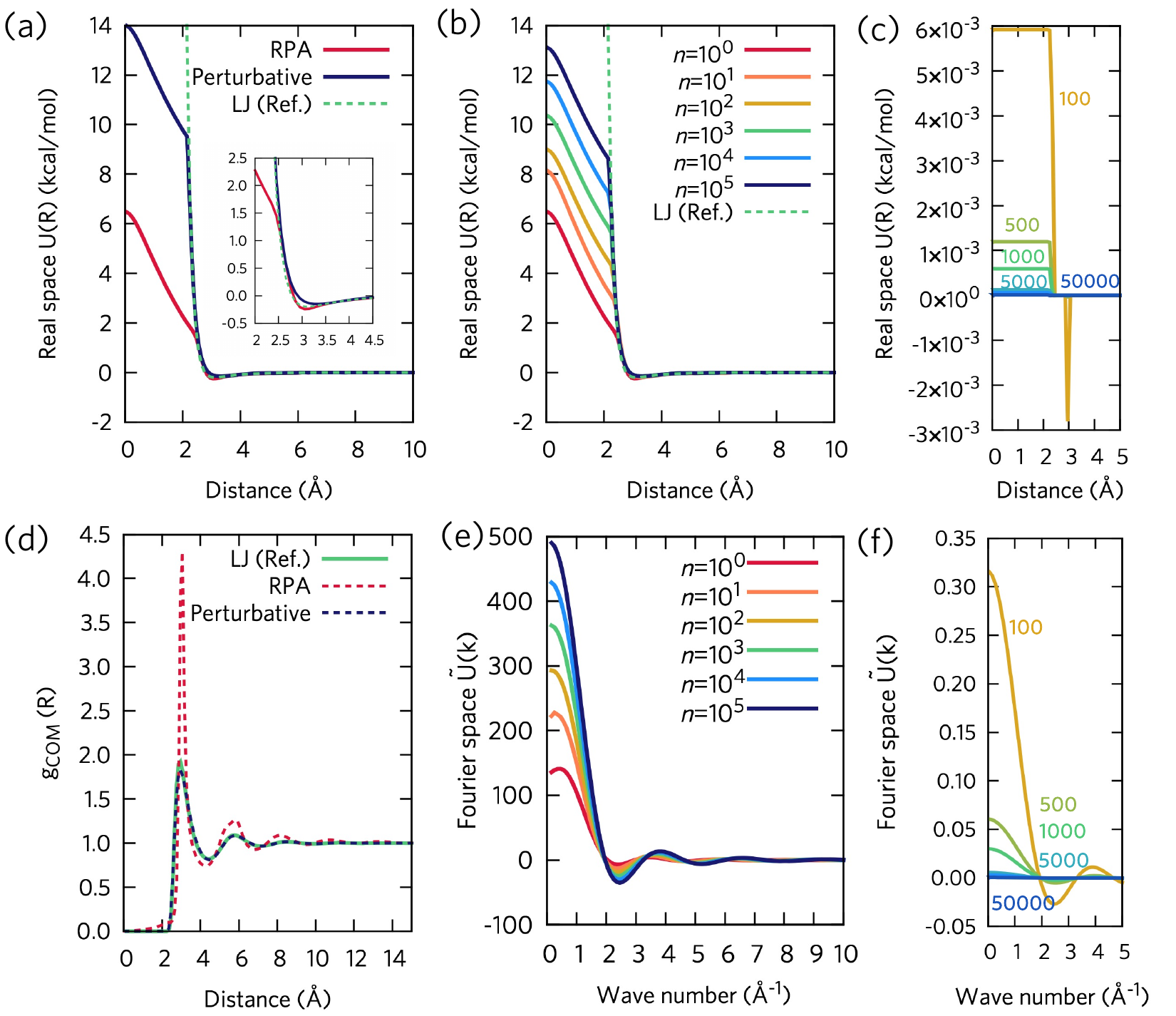"}\\
\textbf{Figure 4}: Perturbative expansion of the hard-core Lennard-Jones liquid ($T^\ast = 3.0$). (a) Overview of interactions in real space where the corresponding RDFs are shown in (d). The truncated perturbative interactions up to the $n$-th order are shown for (b) the real space representation and (e) the Fourier space representation. Each order of the perturbative expansion is depicted in (c) real space (square-like) and (d) Fourier space (Bessel function-like).
\end{center}

From Fig. 4, we conclude that the higher-order perturbative interactions can effectively correct the mismatch observed in the RPA. While lower-order perturbations can partially correct the long-range regime, for larger $n$ values, the perturbation behaves as a square-like well in real space, resembling a hard-core interaction behavior within a distance of 2 \angstrom. As the $n$ value increases, the behavior of these perturbative interactions aligns with the convergence analysis described in Eq. (\ref{eq:convergence}). 


Nevertheless, we can truncate the higher-order contributions after a certain value of $n$. To systematically determine this truncation order, we estimate the overall potential energy using $\tilde{U}^{(n)}$ in $k$-space by computing the following quantity:
\begin{equation}
    U_\mathrm{approx}(\mathbf{R}^N) = \frac{1}{2}\sum_k \tilde{\rho}(k) \tilde{U}^{(n)}(k)\tilde{\rho}(k) -U^{(n)}_\mathrm{self}.
    \label{eq:approxval}
\end{equation}
Equation (\ref{eq:approxval}) was numerically computed using the sampled $\tilde{\rho}(k)$ from MD simulation, and $\mathcal{F}^{-1}[\tilde{U}^{(n)}](0)$ was used to estimate the approximated self-interaction. Unlike the bare LJ interaction, $U_\mathrm{self}^{(n)}$ can be well-defined and computed analytically with our perturbative treatment. If approximated correctly, $U_\mathrm{approx}(\mathbf{R}^N)$ should be identical to the reference potential energy using the interaction $U^\mathrm{HNC}(R)$ estimated by directly solving the HNC closure: $0.5\sum_k \tilde{\rho}(k) \tilde{U}^\mathrm{HNC}\tilde{\rho}(k) -U^{\mathrm{HNC}}_\mathrm{self}$. Figure 5 illustrates the convergence behavior of the total potential energy at varying perturbation order $n$. Similar to the trend shown in Fig. 4(b), the inaccurate potential energy predicted by lower orders is then corrected by including higher-order terms. Following this analysis, hereafter, we will use the same procedure to determine the perturbation orders for other systems.

\begin{center}
\includegraphics[scale=0.35]{"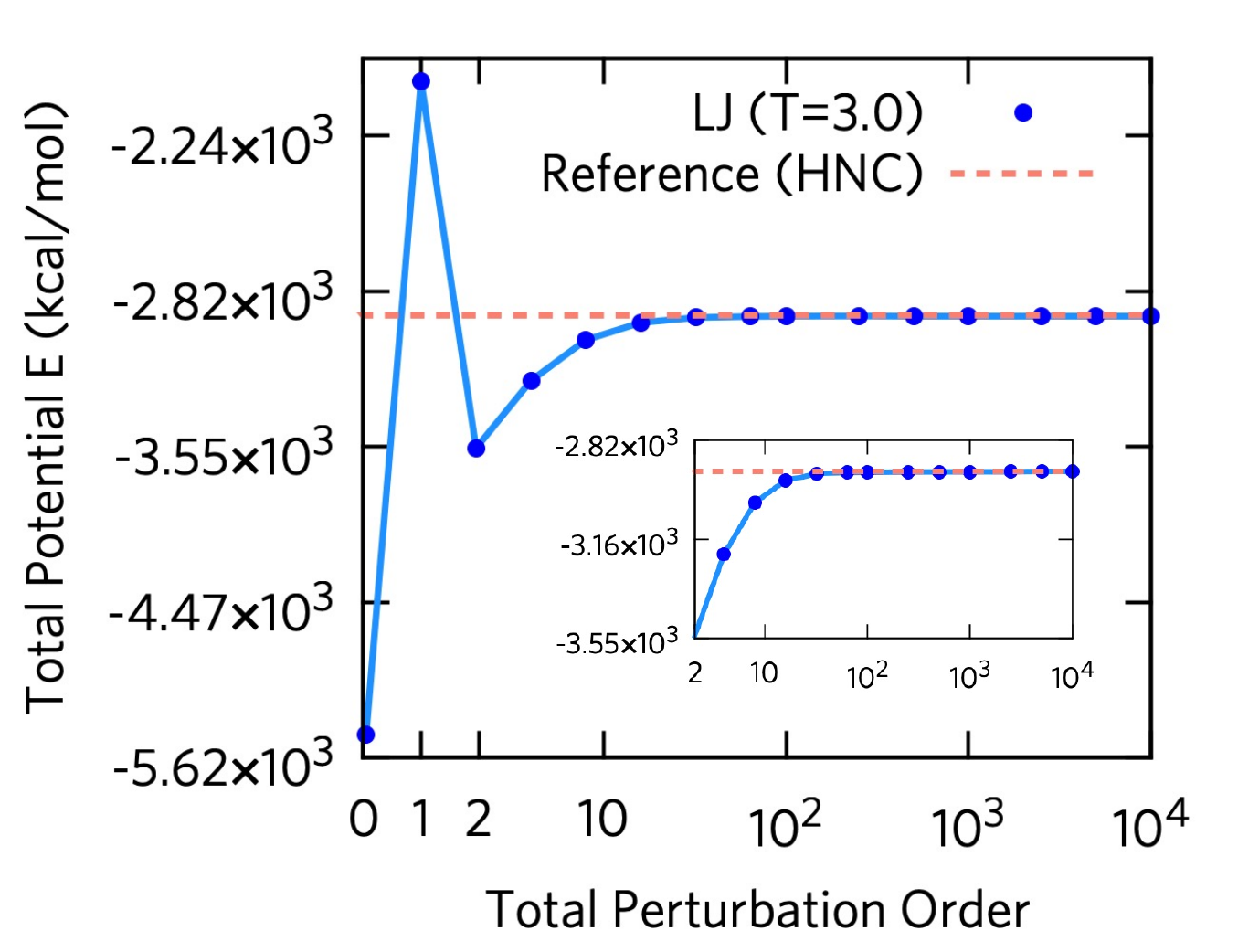"}\\
\textbf{Figure 5}: Determination of the truncation order for the Lennard-Jones liquids at $T^\ast=3.0$. The $n$-th order perturbative expansion (blue dots) approaches to the reference potential energy from the HNC closure at higher orders (red dashed).
\end{center}

We would like to emphasize that our approach maintains a perturbative structure and allows for systematic truncation of higher-order interaction terms. This stands in contrast to conventional \textit{ad hoc} treatments that manually cut off the interaction after a certain value. Figure 4(b) clearly manifests the major differences between our approach and \textit{ad hoc} truncation. In the correct truncated representation of hard-core interaction, the interaction profile should look like a slanted triangle due to the contribution from the RPA term, rather than a square-like flat well. The final approximated LJ interaction truncated at $n=5 \cdot 10^5$ is depicted in Fig. 4(b) and 4(e). As shown in Fig. 5, this interaction reproduces the potential energy estimated by the HNC closure. A slight discrepancy is observed at the first coordination shell and can be attributed to the HNC closure's limitations in accurately describing LJ systems.\cite{levesque1966etude} We also applied our method to the lower temperature LJ system ($T^\ast=2.5$) and observed similar trends. A detailed illustration of these results can be found in Appendix {B}. To go beyond the HNC closure and adequately capture the structural correlations in Fourier space, a more systematic approach is required and will be discussed in {Sec. \ref{sec:bridge}}.  

\subsection{Complex Molecular Systems} \label{sec:CG}

\subsubsection{Molecular Coarse-Graining}

A systematic extension of our approach to complex and realistic systems is crucial in order to develop a bottom-up field-theoretical model that precisely contains the microscopic origins of the system. In the context of field-theoretic simulations, developing a faithful field representation of complex atomistic molecules rather than relying on simple analytical models can be advantageous for efficiently exploring behavior beyond that of current conventional particle simulations. Therefore, as a first step in this direction, the goal of this subsection is to extend the current approach to molecular liquids. 

Unlike the simple models studied in earlier sections, constructing a faithful classical field and its corresponding Fourier representation for molecular systems poses practical challenges for several reasons. Firstly, realistic liquids exhibit a hard-core (divergent) nature, as can be understood from classical perturbation theory. Moreover, at the atomistic level, molecules exhibit a myriad of interactions ranging from non-bonding van der Waals interactions to hydrogen bonding, making the direct construction of a field-theoretical description at the atomistic level highly unfeasible. In particular, various interactions exerted by different atoms within molecules will exacerbate this problem. Nevertheless, the aforementioned approach can be readily translated to the center-of-mass level of atomistic liquids, a procedure known as coarse-graining,\cite{noid2013perspective,saunders2013coarse,jin2022bottom,noid2023perspective} which involves representing complex molecules in a simplified (center-of-mass) representation. Under this {CG} representation, the hard-core nature of molecules would be an ideal target for our approach. 

While physically important, it is a highly challenging task to carry out this problem as the exact analytical form of the underlying interactions is unknown, unlike in molecular mechanics force fields used in full atomistic descriptions.\cite{allinger1989molecular,mayo1990dreiding,rappe1992uff,scott1999gromos,case2005amber,brooks2009charmm} Since the effective interaction of {CG} systems can be understood as the renormalized atomistic interactions obtained by integrating out unnecessary degrees of freedom, this formal interaction is represented as a high-dimensional quantity (often referred to as the many-body potential of mean forces) that cannot be determined exactly.\cite{noid2008multiscale1,noid2008multiscale2} Yet, despite these challenges, our approach can be faithfully extended to the regime of molecular coarse-graining by inferring the center-of-mass level interaction by directly solving the OZ equation with the HNC closure. 

Our strategy for molecular liquids can be described in detail as follows: We begin with a fully atomistic simulation and manually coarse-grain each molecule to its center-of-mass, resulting in configurations represented by the configuration $R$. We can estimate the effective pairwise CG interaction $U_{CG}(R)$ by solving $    h(R)=c(R)+\rho \int c(|R-R'|)h(R')dR'$, where $h(R)=g(R)-1$ can be calculated from the mapped atomistic trajectory. Then, we apply our perturbation theory to the {CG} representation in a manner similar to the approach presented in Sec. {\ref{sec:CG}}. This allows us to estimate the approximated Fourier space representation of $U_{CG}(R)$ for CG systems. Importantly, as our theory is general enough to be applicable to a wide range of interactions, it does not rely on specific forms of interaction potentials or correlations.

\subsubsection{System Settings and Computational Details}

In order to consider a diverse range of intermolecular interactions, we selected three representative molecular liquids that exhibit either van der Waals, polar protic, or hydrogen bonding interactions in the respective order: carbon tetrachloride, methanol, and water. Atomistic simulations were conducted for these liquids using a simulation protocol reported in the {CG} literature at the ambient condition, where the OZ equation remains valid.  

For carbon tetrachloride and methanol, we initially created liquid systems comprised of 1000 molecules in a cubic box. The box lengths were determined based on equilibrium conditions at 300 K and 1 atm, as reported in the literature for carbon tetrachloride\cite{jin2018ultra} and methanol.\cite{jin2018ultra,jin2020temperature} The OPLS-AA force field\cite{jorgensen1996development,kaminski2001evaluation} was employed to describe the atomistic energetics with long-range electrostatics computed using Ewald summation. We then performed constant NVT dynamics to the equilibrated configuration at 300 K using a N\'{o}se-Hoover thermostat for 5 ns with $\tau_{NVT}=0.1 \textrm{ps}$. During the final 2.5 ns of the simulation, configurations were sampled every 0.5 ps for subsequent analysis.  

For water, the initial system consisted of 8000 molecules in a cubic box with a box length of 62.555 \angstrom. This box dimension corresponds to the equilibrium density condition at 300 K and 1 atm.{\cite{jin2021new1, jin2021new2}} We utilized the SPC/Fw force field,\cite{wu2006flexible} even though other water force fields could be in principle implemented without loss of generality. Similar to the simulations for carbon tetrachloride and methanol, we conducted constant NVT dynamics for the water system with the identical protocol.

\subsubsection{Carbon tetrachloride}

Due to the $T_d$ symmetry, it is reasonable to approximate the carbon tetrachloride molecule as a spherical object through center-of-mass coarse-graining. The reference CG interaction was estimated by inverting the OZ equation, as shown in Fig. 6(a). It should be noted that, similar to the LJ system, inverting the OZ equation can only be done in the regime where $g(R)\neq 0$ due to the hard-core nature. However, this short-range repulsion is often extrapolated using a divergent functional form of $A \cdot R^{-B}$.\cite{lu2010efficient} Our particular objective is then to determine if our approach can faithfully capture this highly repulsive behavior. 

As expected, we note that the zeroth-order RPA closure is able to capture the interaction profile of carbon tetrachloride reasonably well, including the interaction minimum within 1 $\angstrom$ [Fig. 6(a)]. However, at short ranges, the RPA closure deviates, and results in a finite repulsion value at zero distance, leading to a non-zero $g(R)$ value in Fig. 6(d). Figure 6(d) was obtained by performing MD simulations using this interaction. This deviation can be ameliorated by including higher-order perturbation terms, where each term clearly exhibits a rectangular-like profile in real space [Fig. 6(c)] and a Bessel-like profile in Fourier space [Fig. 6(f)].

By incorporating these perturbative interactions, our final approximated Fourier representation [Fig. 6(e)] of the real space interaction [Fig. 6(d)] significantly improves the interaction profile when compared to the OZ-CG reference. This agreement is further supported by the RDF of the perturbative simulations, where the short-range non-zero behavior at 2-4 $\angstrom$ is corrected [Fig. 6(d)] as well as capturing the correct first peak value. Overall, Fig. 6 demonstrates the effectiveness of our approach in dealing with more complex molecules. 

\begin{center}
\includegraphics[scale=0.5]{"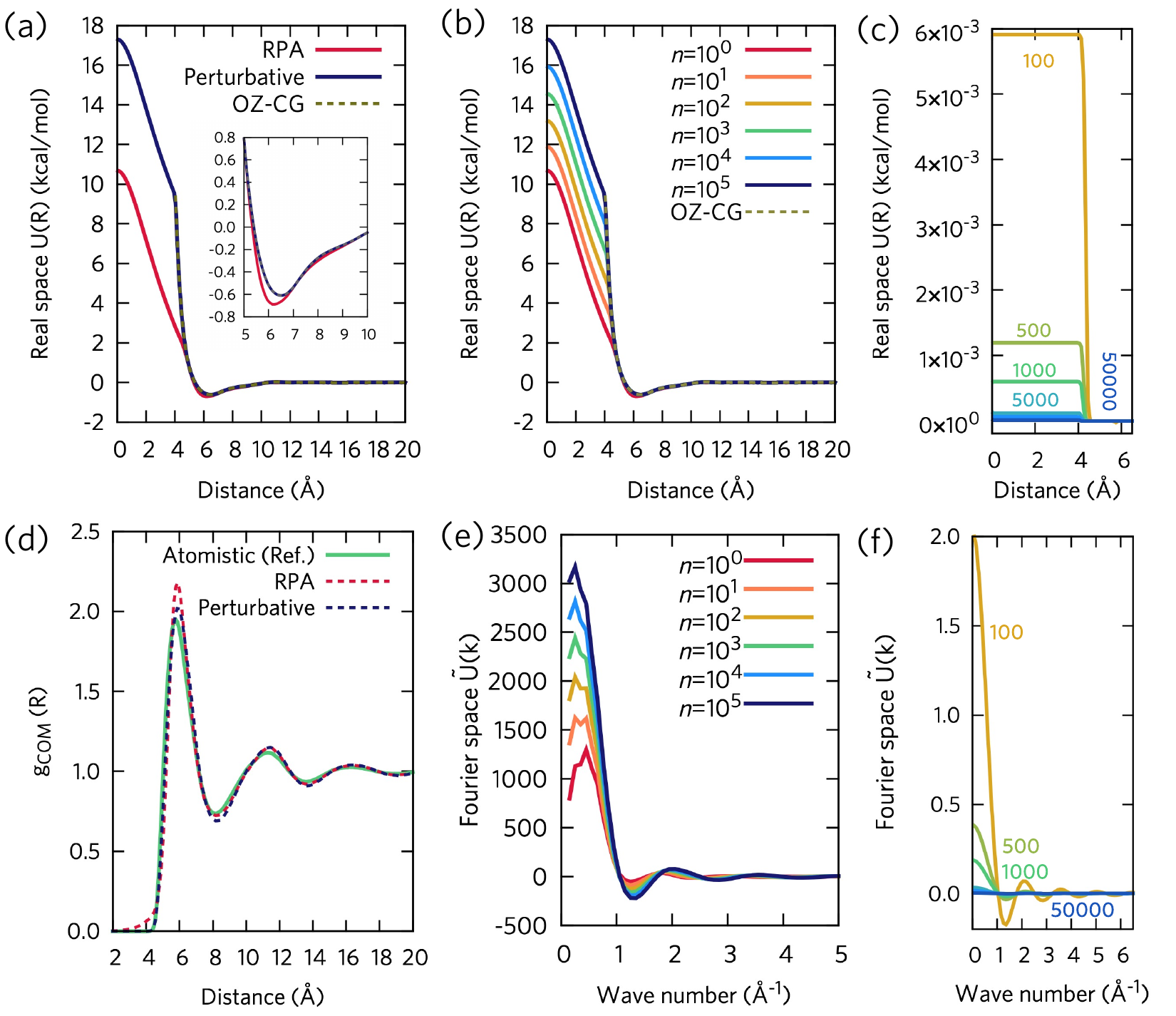"}\\
\textbf{Figure 6}: Perturbative expansion of the coarse-grained carbon tetrachloride (single-site). (a) Overview of interactions in real space where the corresponding RDFs are shown in (d). The truncated perturbative interactions up to the $n$-th order are shown for (b) the real space representation and (e) the Fourier space representation. Each order of the perturbative expansion is depicted in (c) real space (square-like) and (d) Fourier space (Bessel function-like).
\end{center}

\subsubsection{Methanol}

Next, we proceeded to investigate a non-spherical polar molecule, methanol, at the single-site level to validate our methodology. Due to its rod-like shape, the center-of-mass representation of pair CG interactions, as inferred by the OZ equation [Fig. 7(a)], exhibits double wells,\cite{golubkov2006generalized} which is distinct from the behavior observed in the aforementioned systems. While the RPA is able to capture the long-range behavior reasonably well, including the first two interaction wells, it produces a relatively smaller value for $U^{(0)}(R=0)$, resulting in a pair correlation that deviates significantly from the atomistic reference [Fig. 7(d)]. This significant discrepancy suggests that the RPA closure may not be suitable for non-spherical systems with complex interaction profiles. 

By incorporating higher-order perturbative interactions in Fourier space [Fig. 7(e)], the corresponding real space MD simulation using Fig. 7(b) validates that the solid-like behavior observed in the zeroth-order perturbation can be enhanced by including higher-order perturbation terms. Both $\tilde{U}^{(n)}$ and $U^{(n)}$ exhibit similar interaction profiles as observed in the methanol or LJ system, further confirming the effectiveness of considering higher-order perturbations in improving the system's behavior.

\begin{center}
\includegraphics[scale=0.5]{"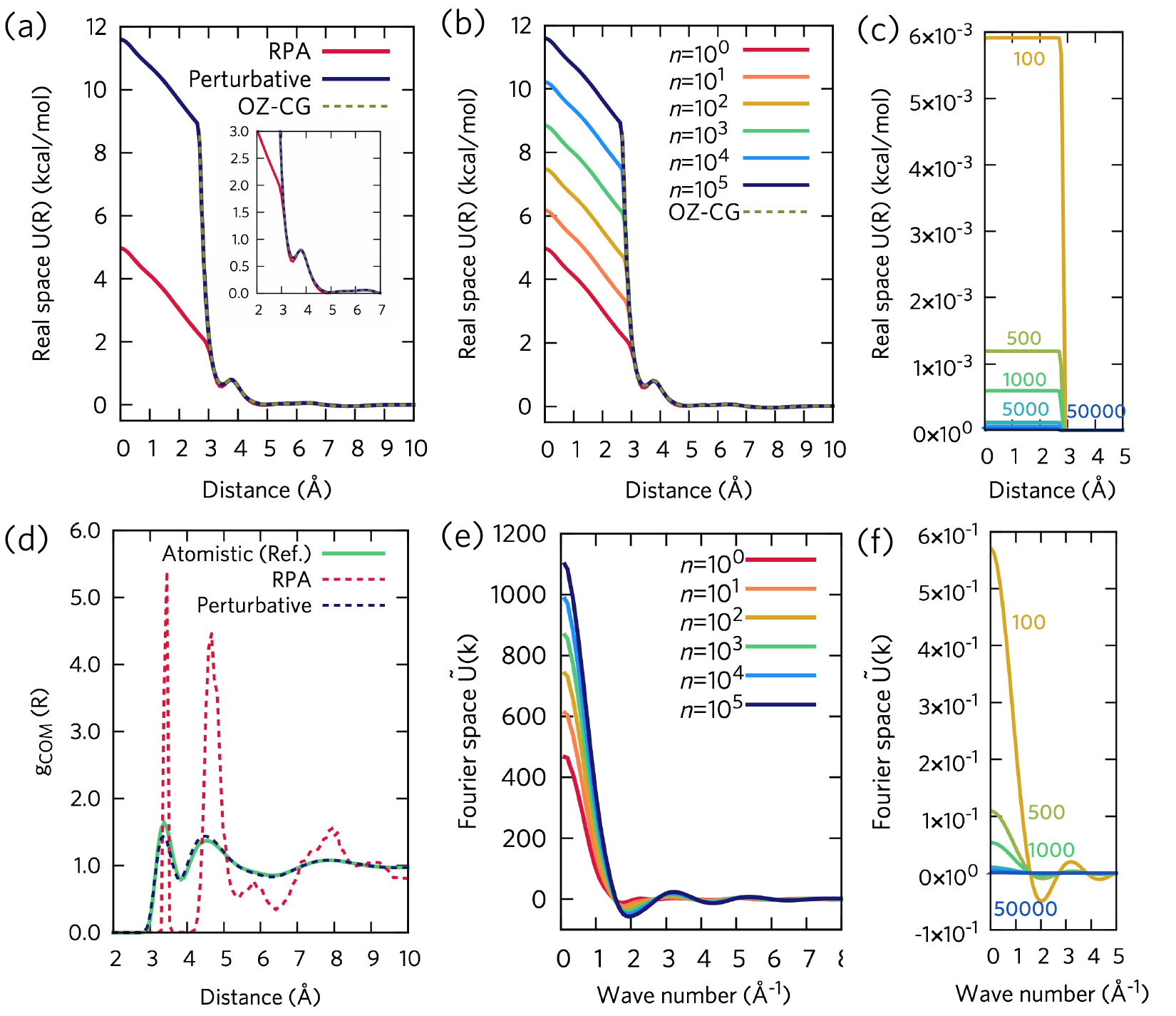"}\\
\textbf{Figure 7}: Perturbative expansion of the coarse-grained methanol (single-site). (a) Overview of interactions in real space where the corresponding RDFs are shown in (d). The truncated perturbative interactions up to the $n$-th order are shown for (b) the real space representation and (e) the Fourier space representation. Each order of the perturbative expansion is depicted in (c) real space (square-like) and (d) Fourier space (Bessel function-like).
\end{center}

\subsubsection{Water} \label{subsec:water}

Finally, we apply our theory to liquid water. Liquid water is of significant importance in the development of a bottom-up field theoretical model due to its ubiquitous presence in various systems.\cite{tait1971water,israelachvili1996role,dill2005modeling} In order to investigate inhomogeneous interfaces or solvated biological systems, it becomes necessary to consider the impact of solvents explicitly. However, employing brute force MD simulations for such large systems becomes computationally prohibitive when exploring long-time scale processes. In this regard, the field-theoretical approach, which treats multiple solvent molecules as a single field, offers a promising solution to greatly enhance the computational efficiency of computer simulations.

Another notable characteristic of water arises from hydrogen bonding,\cite{davidson1973water,eisenberg2005structure} which manifests from higher-order interactions as reported in the literature. However, in this study, we limit our description of {CG} water to effective two-body interactions. Nevertheless, it is worth noting that the approach presented here can be extended to incorporate many-body interactions using the same underlying principles.\cite{hankins1970water}

Figure 8 succinctly summarizes our findings for liquid water. The CG interaction of water obtained from the OZ equation exhibits a shouldered double-well shape, which aligns with other pairwise CG water interactions derived through bottom-up approaches. {Unlike the previous systems, liquid water exhibits stronger correlations with a $g(R)$ of approximately 2.9 at 2.75 \angstrom. Therefore, the perturbative structure derived from Eq. (\ref{eq:Taylor}) cannot be applied here. Instead, using Eq. (\ref{eq:m-expansion}), we can derive the following perturbative expansion in reciprocal space:
\begin{equation}
    c(R) = \left[ \left(1-\frac{1}{m}\right)g(R)-\ln m\right] + \frac{(h(R)-m+1)^2}{2m^2} - \frac{(h(R)-m+1)^3}{3m^3} + \mathcal{O}[h(R)^4] - \beta U(R).
    \label{eq:highdensity-cR}
\end{equation}
From Eq. (\ref{eq:highdensity-cR}) with $m=\lfloor g(R) \rfloor =2$, we arrive at the perturbative expansion with higher-order corrections for water:
\begin{equation}
    U(R) = -\frac{c(R)}{\beta} +\frac{1}{\beta} \left[ \left( \frac{g(R)}{2}-\ln 2 \right) + \frac{1}{2} \left(\frac{h(R)-1}{2} \right)^2 + \frac{1}{3} \left(\frac{h(R)-1}{2} \right)^3 + \cdots\right]
\end{equation}
It is worth noting that for systems with stronger correlations ($m>1$), the zeroth-order truncation of the HNC closure no longer corresponds to the RPA closure. For the zeroth-order truncation, this additional contribution to the RPA slightly increases $U(R)$ near the local potential energy minimum (2.5--3 \angstrom). While} the zeroth-order RPA deviates significantly from the reference structure, {this overly structured profile can still be improved at the zeroth-order by including this additional contribution [Fig. 8(a)].} By including perturbation terms up to an order of $n=10^4$, the approximated perturbative interaction closely matches the reference CG interaction, accurately capturing the $g(R)$ at the center-of-mass level. In summary, the three molecular systems presented in this section underscore the versatility of our approach that takes advantage of bottom-up molecular coarse-graining.

\begin{center}
\includegraphics[scale=0.4]{"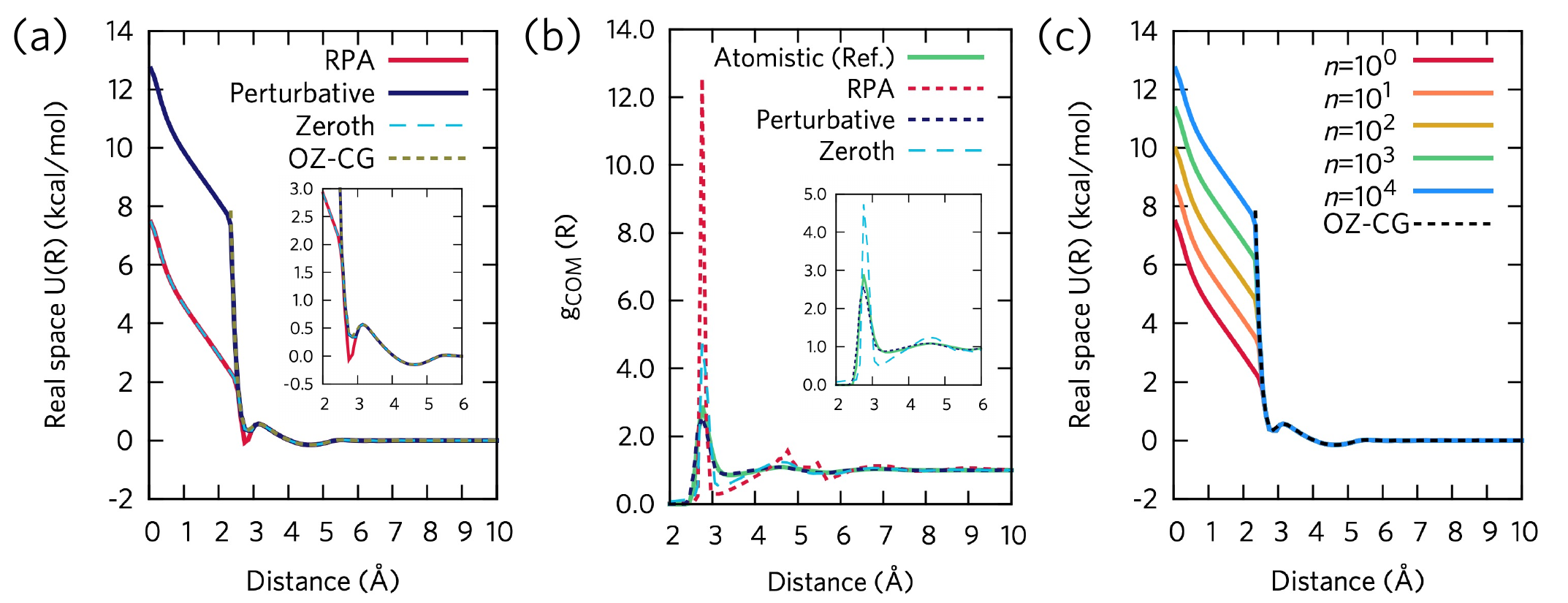"}\\
\textbf{Figure 8}: Perturbative expansion of the coarse-grained water (single-site) using the SPC/Fw model.  (a) Overview of interactions in real space where the corresponding RDFs are shown in (b). The truncated perturbative interactions up to the $n$-th order are shown for (c) the real space representation. {For high-density systems [$g(R) > 2$], the zeroth-order perturbation no longer follows the RPA form but can correct the overstructured peak from the RPA}.
\end{center}

\subsection{A Microscopically-Inferred Bridge Function: Beyond the HNC Closure} \label{sec:bridge}

\subsubsection{Importance of the Bridge Function}

Up to this point, we have exclusively considered the RPA and HNC closures when approximating the interactions in real and reciprocal spaces by closing the OZ equation. As demonstrated earlier, the RPA closure tends to deviate for complex interactions, while the HNC closure generally provides a reasonable description for liquids. Yet, it is important to note that the HNC closure itself is an approximation, as observed in the LJ system and certain molecular liquids. This is because the HNC closure assumes a zero bridge function $B(R) = 0$, and the approximated pair interaction is expressed as  
\begin{equation}
    \beta U^\ast(R)=-\ln g^\ast(R)+g^\ast(R)-1-c^\ast(R).
    \label{eq:ApproxBridge}
\end{equation}
Here, the superscript $\ast$ denotes that the pairwise functions in Eq. (\ref{eq:ApproxBridge}) are approximations to the reference value using the HNC approximation. Hence, incorporating the non-zero bridge function, derived from microscopic information, could potentially extend the range and accuracy of our proposed method beyond simple closure relationships.\cite{morita1960new,duh1996integral} In principle, the effect of the bridge function should be taken into account as 
\begin{equation}
    \beta U(R)=-\ln g(R)+g(R)-1-c(R)+B(R),
    \label{eq:bridge}
\end{equation}
but an exact determination of the bridge function for a given system requires an infinite series of diagrammatic summations, making it practically infeasible and limited in practice. 

\subsubsection{Bottom-up Parameterization of Bridge Function}

Given the bottom-up nature of our coarse-graining approach, we propose a bottom-up strategy to estimate the bridge function numerically from the microscopic statistics. For model systems such as the GCM and LJ potentials, where the exact form of $U(R)$ and the corresponding microscopic statistics $g(R)$, as well as $c(R)$, are known from MD simulations, we can directly infer $B(R)$ by inverting Eq. (\ref{eq:bridge}) as $B(R) \approx \beta U(R)+\ln g(R)-g(R)+1+c(R)$. Note that we use $\approx$ here because we numerically estimated the approximate correlation functions using MD simulation. 

For molecular systems, unlike model systems, there is no reference analytical potential readily available, and it is often ambiguous to define the effective interaction potential $U(R)$. In such cases, bottom-up molecular coarse-graining approaches come into play.\cite{izvekov2005multiscale1,izvekov2005multiscale2,noid2008multiscale1,noid2008multiscale2} The main objective of bottom-up coarse-graining is to determine effective equations of motion in terms of {CG} variables from atomistic simulations. Detailed surveys and reviews on coarse-graining can be found in Ref. \citenum{noid2013perspective,saunders2013coarse,jin2022bottom,noid2023perspective}. In our study, we focus on determining the effective CG interactions at the center-of-mass level by utilizing only atomistic statistics. Among various methods available, such as iterative Boltzmann inversion (IBI)\cite{reith2003deriving} or multiscale coarse-graining (MS-CG),\cite{izvekov2005multiscale1,izvekov2005multiscale2,noid2008multiscale1,noid2008multiscale2} we employ the relative entropy minimization (REM) approach.\cite{shell2008relative,chaimovich2010relative,chaimovich2011coarse,shell2016coarse} Essentially, this approach aims to minimize the Kullback-Leibler divergence between the target (atomistic) and CG distributions. It has been demonstrated that the REM method accurately captures pairwise statistics,\cite{chaimovich2010relative} which are crucial for our work due to the presence of $g(R)$ and $c(R)$ in Eq. (\ref{eq:bridge}). {Furthermore, we note that similar principles can be applied to top-down approaches, as introduced in Sec. \ref{subsec:topdown}. Here, the Empirical Potential Structure Refinement (EPSR)\cite{soper1996empirical} can be leveraged to refine the interaction potential $U(R)$ through an iterative inverse Monte Carlo method.\cite{lyubartsev1995calculation} By assuming that the experimental correlation perfectly encodes the effects of the bridge function, $B(R)$ can be numerically estimated from the updated $U(R)$ via a top-down approach as well.}

In practice, the {CG} interaction $U_{CG}(R)$ was variationally determined by minimizing the following functional, which is commonly referred to as the Kullback-Leibler divergence or the relative entropy $S_\mathrm{rel}[U]$,\cite{kullback1951information}
\begin{equation}
    S_\mathrm{rel}[U_{CG}] = \langle S_\mathrm{map} \rangle + \int d\mathbf{R}^N p_\mathrm{ref} (\mathbf{R}^N) \ln \frac{p_\mathrm{ref} (\mathbf{R}^N)}{p (\mathbf{R}^N|U_{CG})}.
    \label{eq:rem}
\end{equation}

In Eq. (\ref{eq:rem}), the mapping entropy originates from the CG mapping operator, i.e., $\langle S_\mathrm{map} \rangle = \ln \int d\mathbf{r}^n \delta[\mathbf{M}(\mathbf{r}^n)-\mathbf{R}^N]$, and $p_\mathrm{ref}(\mathbf{R}^N)$, $p(\mathbf{R}^N|U_{CG})$ correspond to the reference (center-of-mass level) and CG probability distribution function, respectively, determined by the CG potential $U_{CG}$. In the canonical ensemble, Eq. (\ref{eq:rem}) can be written in a more tractable form as
\begin{equation}
    S_\mathrm{rel}[U_{CG}] = [\beta \langle E_{FG} - E_{CG} \rangle_{CG} - (A_{FG}-A_{CG})] + \langle S_\mathrm{map} \rangle,
    \label{eq:rem2}
\end{equation}
where $E$ and $A$ are the internal energy and free energy {of the fine-grained (FG) and CG systems}.\cite{chaimovich2011coarse} Now, from Eq. (\ref{eq:rem2}), a pairwise CG potential represented using spline basis sets $\{\lambda_i\}$ can be variationally optimized by minimizing $S_\mathrm{rel}$ with respect to $\lambda_i$. This is achieved by setting $\partial S_\mathrm{rel}/\partial \lambda_i = 0$ and ensuring that $\partial^2 S_\mathrm{rel}/\partial\lambda_i^2 > 0$ using a Newton-Raphson scheme. For the optimization process, we closely followed the numerical settings reported in Refs. \citenum{shell2008relative,chaimovich2010relative,chaimovich2011coarse} to construct the CG liquid models using the open source program described in Ref. \citenum{OpenMSCG}.

Once $U_{CG}(R)$ is determined through the minimization of $S_\mathrm{rel}$, we performed additional MD simulations utilizing the optimized CG interactions. The initial configuration for the MD run was generated by mapping the molecules to their center-of-mass configurations based on the last snapshot of the atomistic simulations. The same thermostat settings as the atomistic simulation were applied to perform constant \textit{NVT} simulations for 5 ns. From the last 2.5 ns of the trajectory, we numerically estimated the correlation functions $g_\mathrm{COM}(R)$ and $c_\mathrm{COM}(R)$ based on the center-of-mass. Finally, the bridge function can be microscopically inferred as
\begin{equation}
    B_{CG}(R) \approx \beta U_{CG}(R)+\ln g_\mathrm{COM}(R)-g_\mathrm{COM}(R)+1+c_\mathrm{COM}(R).
    \label{eq:bridge2}
\end{equation}

Since $B_{CG}(R)$ obtained through Eq.(\ref{eq:bridge2}) is finite (i.e., we only sample when $g_\mathrm{COM}>0$), it can be seamlessly incorporated into the perturbative treatment presented in the previous sections. In particular, considering that $B(R) = 0$ for the HNC closure, the non-zero contribution from $B_{CG}(R)$ determined by Eq. (\ref{eq:bridge2}) can be added to Eq. (\ref{eq:R-Taylor}). Finally, we arrive at the following expressions for the potential energy in real space and reciprocal space: 
\begin{equation}
    U(R) = \frac{B_{CG}(R)-c(R)}{\beta} +\frac{1}{\beta}\left( \frac{h^2(R)}{2} - \frac{h^3(R)}{3}+\cdots \right),
    \label{eq:BR-Taylor}
\end{equation}
and 
\begin{equation}
    \tilde{U}(k)=\frac{1}{\beta}\left[\frac{1}{\rho } \left(\frac{1}{S(k)}-1\right)+\mathcal{F}[B_{CG}(R)] \right]+\frac{1}{\beta}\left(\frac{\tilde{h}^2(k)}{2}-\frac{\tilde{h}^3(k)}{3} +\cdots \right),
    \label{eq:Bk-Taylor}
\end{equation}
respectively. Equation (\ref{eq:Bk-Taylor}) presents an improved representation of the Fourier transformed $\tilde{U}(k)$ by effectively integrating the microscopically-inferred bridge function. This equation also underscores the key distinction of our approach from the mode expansion.\cite{andersen1970mode,chandler1971mode,andersen1971mode} 
While both approaches are rooted in RPA-based perturbations, beyond the RPA description, the mode expansion employs the Mayer-like cluster expansion\cite{mayer1940statistical} to account for higher-order correlations. Conversely, in our case, the higher-order perturbations are derived from expanding the HNC closure [$B(R)=0$], and the effect of a non-zero bridge function is then described by a single term inferred from microscopic statistics.

\subsubsection{Results}

To verify the accuracy of our microscopically-inferred bridge function, we first applied Eqs. (\ref{eq:bridge}) and (\ref{eq:bridge2}) to the LJ system. Figure 9(a) illustrates the scaled quantity $k_BT \cdot B_{LJ}(R)$ estimated using Eq. (\ref{eq:bridge}) for the $T^\ast = 3.0$ condition. Note that we only sampled $B_{LJ}(R)$ down to 2.2 \angstrom, where $g(R) > 0$. The inferred bridge function shows a short-ranged feature\cite{rosenfeld1979theory} that begins to decay below $R=\sigma $ with a concave curvature, which aligns with the previous studies that determined $B_{LJ}(R)$ from the cavity function formalism.\cite{llano1992bridge}

Next, we incorporated the non-zero influence of $B_{LJ}(R)$ into the perturbative expression for $\mathcal{F}^{-1}[\tilde{U}^{(n)}]$ and performed MD simulations. The resulting RDF is depicted in Fig. 9(b), which demonstrates a noticeable enhancement of the first peak. As illustrated in Fig. 4(d), the HNC closure previously underestimated the intensity of the first peak by 0.2 at 3 \angstrom. However, with the inclusion of the inferred $B_{LJ}(R)$, this discrepancy is alleviated, leading to an improved agreement with the reference data.

\begin{center}
\includegraphics[scale=0.125]{"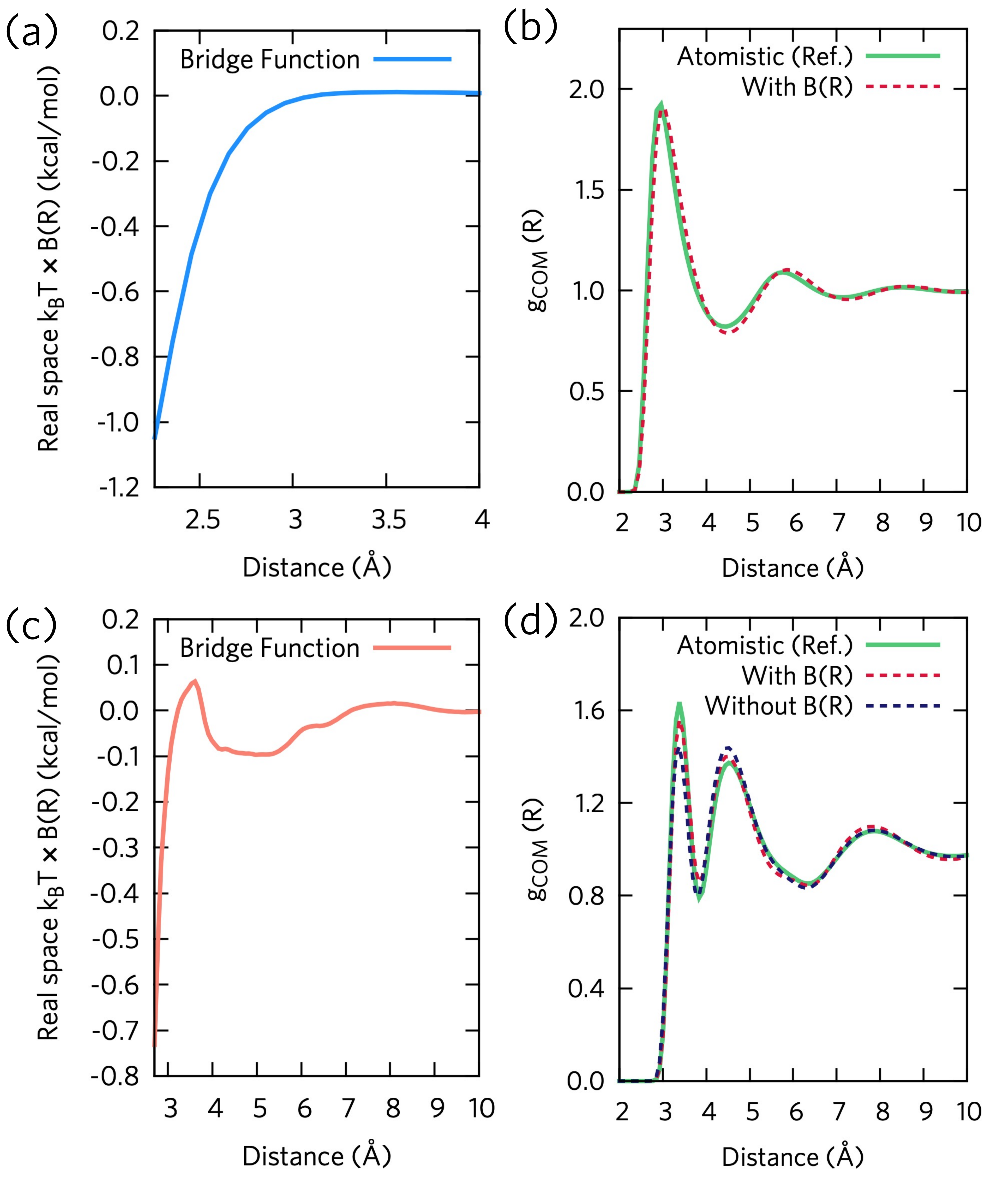"}\\
\textbf{Figure 9}: Perturbative approach beyond the HNC approximation by inferring the bridge functions for complex molecules: (a-b) Lennard-Jones liquid as a proof-of-concept and (c-d) methanol. Panels (a) and (c) display the bridge functions parametrized by microscopic coarse-graining. Panels (b) and (d) demonstrate the enhanced center-of-mass RDFs achieved by incorporating $B(R)$. 
\end{center}

To further demonstrate the applicability and accuracy of our microscopic approach to more complex systems with non-trivial interactions, we extended our analysis to the CG methanol system. The bottom-up parametrized CG interaction for liquid methanol was determined by employing the relative entropy minimization as described in the previous subsection. The resulting interaction profile exhibits similarities to previous studies on CG methanol. By utilizing $U_{CG}(R)$ and the center-of-mass level correlation functions, $B_{CG}(R)$ was then estimated using Eq. (\ref{eq:bridge2}). The inferred bridge function for methanol [Fig. 9(c)] displays a complex profile in comparison with the LJ [Fig. 9(a)], which can be understood from the non-isotropic nature of the single-site representation of methanol molecules. Yet, $B_{CG}(R)$ also exhibits a similar decaying behavior at short distances when $R\rightarrow 0$. Remarkably, with the incorporation of $B_{CG}(R)$ into the corrected $\mathcal{F}^{-1}[\tilde{U}(k)](R)$, our model successfully improves the RDF for methanol as well. Particularly, the inclusion of $B_{CG}(R)$ corrects the previously underestimated first peak and es the over-structured second peak in the RDF.

To summarize, this section showcased the successful application of our approach in determining the bridge function from microscopic statistics for both simple model systems (such as LJ) and complex molecular systems such as methanol. By combining this microscopic bridge function with our perturbative expansion based on the HNC closure, our theory can accurately construct the Fourier space representation of complex molecular interactions. This approach goes beyond simple closures and demonstrates broad applicability and extensibility, making it a valuable tool for studying a wide range of molecular systems.

\section{Conclusions}
In this paper, we have presented a novel statistical mechanical theory for constructing the Fourier space representation of various molecular interactions. Unlike for soft materials with finite repulsions, constructing Fourier representations of interparticle interactions for molecules is an ill-defined problem due to the non-penetrable nature of molecules. Inspired by classical perturbation theory, where long-range attractions are treated as a perturbation in real space, our approach aims to decompose short-range repulsions as a perturbation in reciprocal space. Starting from the Ornstein-Zernike integral equation for liquids, we derived that the zeroth-order perturbation corresponds to the random phase approximation by qualitatively capturing the long-range interactions. Then, under the hypernetted chain closure where the influence of bridge functions was initially neglected, the higher-order perturbation terms in the Fourier representation can be understood as an expansion of the total correlation function. At large orders, this total correlation function becomes a Bessel function in reciprocal space, resulting in a square-like repulsion localized at the hard core region in real space and leading to a divergent potential as $R \rightarrow 0$. By correctly decomposing the hard-core repulsion and long-range attractions and accounting for their contributions to the overall energy, our approach imparts a systematic framework for constructing the Fourier representation of hard-core interactions.

Encompassing analytical models, e.g., Gaussian and Lennard-Jones, as well as complex molecular interactions at the coarse-grained level, e.g., molecular liquids, our theory further unravels how to systematically construct Fourier representations of various microscopic interactions. To enhance the accuracy of our method, we extended our method beyond the hypernetted chain closure by numerically inferring the bridge function from microscopic information and correlations, which is otherwise impractical. The combination of systematic coarse-graining approaches with our proposed approach offers a promising avenue for accurately representing molecular systems in Fourier space. This development is particularly relevant and important for multiscale modeling, as it establishes a connection between the microscopic density operator and the mesoscopic field-theoretical description, enabling a comprehensive understanding of complex molecular systems across different scales.

Moreover, having a well-defined Fourier component $\tilde{U}(k)$ allows the potential energy of the system to be expressed as a quadratic function of the Fourier modes of the density field operator. While soft interactions are commonly employed in existing field simulations of polymers,\cite{fredrickson2002field,fredrickson2006equilibrium,fredrickson2023field} these \textit{ad hoc} interactions typically exhibit soft repulsions at shorter distances (high $k$), which may not accurately capture the hard-core repulsion at the microscopic level. Furthermore, current \textit{ad hoc} approaches often rely on phenomenological model systems, which can lead to the loss of detailed microscopic physics. Overcoming this challenge is especially crucial for understanding morphologies and phase diagrams of polymers. \cite{bates2019stability,zhang2021emergence} {Lastly, beyond solely sampling equilibrium structures from free energies, having a well-defined Fourier representation enables the rigorous description of the mesoscopic dynamics of density field operators. One potential extension would be to integrate the current framework into the Dean-Kawasaki\cite{dean1996langevin,kawasaki1998microscopic} or DDFT\cite{marconi1999dynamic,marconi2000dynamic,archer2004dynamical} dynamics, which includes equations of motion for $\rho(\mathbf{R})$ or $\tilde{\rho}(k)$. Currently, the literature lacks a bottom-up implementation of mesoscopic field dynamics for molecules. Therefore,} by directly starting from microscopic statistics, our findings open up promising avenues for the development of bottom-up field-theoretical simulations. Our novel approach can address the limitations of current field-theoretical modeling of polymers, enhancing the accuracy and effectiveness in capturing the underlying physics. Additional work in this direction is currently underway. Altogether, we expect our framework to rigorously bridge the gap between microscopic physics and mesoscopic field-theoretical modeling by way of bottom-up coarse-graining. 

\begin{acknowledgement}

J.J. thanks the Arnold O. Beckman Postdoctoral Fellowship for funding and academic support.
\end{acknowledgement}

\section*{Appendix}

\subsection*{A. High-Density Expansion}
{For systems exhibiting strong structural correlations (i.e., $g(R) \gg 1$), we can alternatively express the divergent behavior of $\ln g(R)$ as follows. Without loss of generality, if the greatest integer value of $g(R)$ is $m$, i.e., $\lfloor g(R) \rfloor = m$, we can express $\ln g(R)$ as
\begin{equation}
    \ln g(R) = \ln \left[ m+h^{(m)}(R)\right] = \ln m + \sum_{n=1}^\infty (-1)^{n+1} \frac{[h^{(m)}(R)]^n}{n\cdot m^n}
    \label{eq:m-expansion}
\end{equation}
where $h^{(m)}(R)=g(R)-m$. Since this expression converges for $|h^{(m)}(R)| < m$, it is suitable for strongly correlated systems. A further demonstration of Eq. (\ref{eq:m-expansion}) is provided in Sec. \ref{subsec:water}.
}
\subsection*{B. Lennard-Jones System ($T^*=2.5$)}

To assess the robustness of our approach across different temperature regimes, we also validated our analysis to the Lennard-Jones system at a lower temperature ($T^\ast = 2.5$) compared to the previous investigation in Sec. {\ref{sec:lj}}. Remarkably, we observed a consistent trend (compared with that shown in Fig. 4) in the interactions and correlations, as depicted in Figure {B}1. Due to the lower temperature condition, we observe slightly stronger structural correlations, as shown in Fig. {B}1(d). Nevertheless, the overall conclusions from the previous section remained unchanged. For example, we observed that the HNC closure still underestimated the attractive part of the Lennard-Jones interaction [Fig. {B}1(a)]. In terms of perturbative interactions, the respective perturbative term yields a square-like well in real space [Fig. {B}1(c)] and a Bessel-like function in reciprocal space  [Fig. {B}1(f)]. In turn, these features enabled an accurate Fourier representation, successfully reproducing the RDF, as exemplified in Fig. {B}1(d).
\begin{center}
\includegraphics[scale=0.4]{"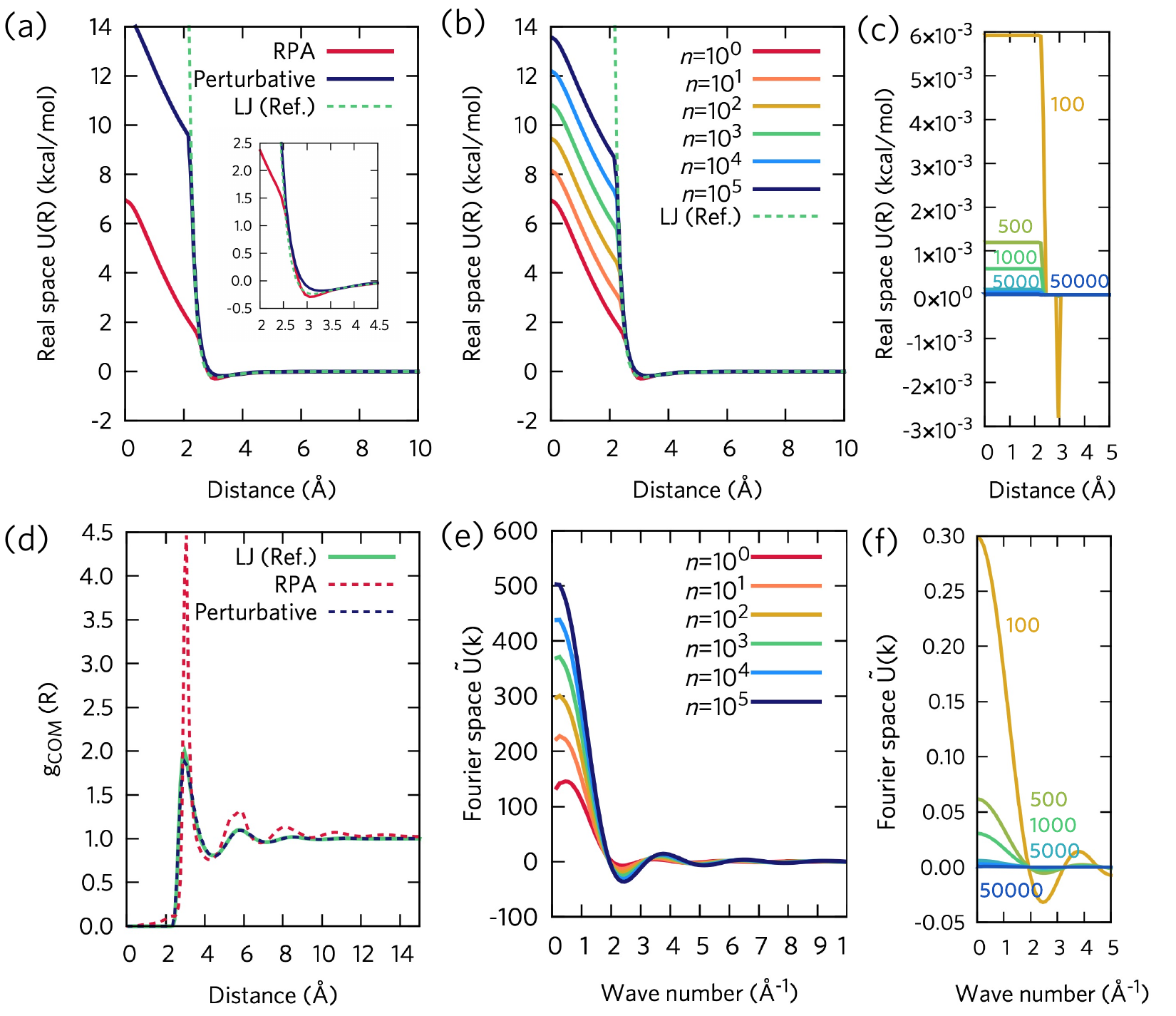"}\\
\textbf{Figure {B}1}: Perturbative expansion of the hard-core Lennard-Jones liquid ($T^\ast = 2.5$). (a) Overview of interactions in real space where the corresponding RDFs are shown in (d). The truncated perturbative interactions up to the $n$-th order are shown for (b) the real space representation and (e) the Fourier space representation. Each order of the perturbative expansion is depicted in (c) real space (square-like) and (d) Fourier space (Bessel function-like).
\end{center}

\bibliography{revised}

\providecommand{\latin}[1]{#1}
\makeatletter
\providecommand{\doi}
  {\begingroup\let\do\@makeother\dospecials
  \catcode`\{=1 \catcode`\}=2 \doi@aux}
\providecommand{\doi@aux}[1]{\endgroup\texttt{#1}}
\makeatother
\providecommand*\mcitethebibliography{\thebibliography}
\csname @ifundefined\endcsname{endmcitethebibliography}
  {\let\endmcitethebibliography\endthebibliography}{}
\begin{mcitethebibliography}{107}
\providecommand*\natexlab[1]{#1}
\providecommand*\mciteSetBstSublistMode[1]{}
\providecommand*\mciteSetBstMaxWidthForm[2]{}
\providecommand*\mciteBstWouldAddEndPuncttrue
  {\def\EndOfBibitem{\unskip.}}
\providecommand*\mciteBstWouldAddEndPunctfalse
  {\let\EndOfBibitem\relax}
\providecommand*\mciteSetBstMidEndSepPunct[3]{}
\providecommand*\mciteSetBstSublistLabelBeginEnd[3]{}
\providecommand*\EndOfBibitem{}
\mciteSetBstSublistMode{f}
\mciteSetBstMaxWidthForm{subitem}{(\alph{mcitesubitemcount})}
\mciteSetBstSublistLabelBeginEnd
  {\mcitemaxwidthsubitemform\space}
  {\relax}
  {\relax}

\bibitem[Frenkel and Smit(2001)Frenkel, and Smit]{frenkel2001understanding}
Frenkel,~D.; Smit,~B. \emph{Understanding molecular simulation: from algorithms
  to applications}; Elsevier, 2001; Vol.~1\relax
\mciteBstWouldAddEndPuncttrue
\mciteSetBstMidEndSepPunct{\mcitedefaultmidpunct}
{\mcitedefaultendpunct}{\mcitedefaultseppunct}\relax
\EndOfBibitem
\bibitem[Allen and Tildesley(2017)Allen, and Tildesley]{allen2017computer}
Allen,~M.~P.; Tildesley,~D.~J. \emph{Computer simulation of liquids}; Oxford
  University Press, 2017\relax
\mciteBstWouldAddEndPuncttrue
\mciteSetBstMidEndSepPunct{\mcitedefaultmidpunct}
{\mcitedefaultendpunct}{\mcitedefaultseppunct}\relax
\EndOfBibitem
\bibitem[Noid(2013)]{noid2013perspective}
Noid,~W.~G. {Perspective: Coarse-grained models for biomolecular systems}.
  \emph{The Journal of Chemical Physics} \textbf{2013}, \emph{139},
  090901\relax
\mciteBstWouldAddEndPuncttrue
\mciteSetBstMidEndSepPunct{\mcitedefaultmidpunct}
{\mcitedefaultendpunct}{\mcitedefaultseppunct}\relax
\EndOfBibitem
\bibitem[Saunders and Voth(2013)Saunders, and Voth]{saunders2013coarse}
Saunders,~M.~G.; Voth,~G.~A. Coarse-graining methods for computational biology.
  \emph{Annual review of biophysics} \textbf{2013}, \emph{42}, 73--93\relax
\mciteBstWouldAddEndPuncttrue
\mciteSetBstMidEndSepPunct{\mcitedefaultmidpunct}
{\mcitedefaultendpunct}{\mcitedefaultseppunct}\relax
\EndOfBibitem
\bibitem[Jin \latin{et~al.}(2022)Jin, Pak, Durumeric, Loose, and
  Voth]{jin2022bottom}
Jin,~J.; Pak,~A.~J.; Durumeric,~A.~E.; Loose,~T.~D.; Voth,~G.~A. Bottom-up
  coarse-graining: Principles and perspectives. \emph{Journal of Chemical
  Theory and Computation} \textbf{2022}, \emph{18}, 5759--5791\relax
\mciteBstWouldAddEndPuncttrue
\mciteSetBstMidEndSepPunct{\mcitedefaultmidpunct}
{\mcitedefaultendpunct}{\mcitedefaultseppunct}\relax
\EndOfBibitem
\bibitem[Noid(2023)]{noid2023perspective}
Noid,~W.~G. Perspective: Advances, Challenges, and Insight for Predictive
  Coarse-Grained Models. \emph{The Journal of Physical Chemistry B}
  \textbf{2023}, \emph{127}, 4174--4207\relax
\mciteBstWouldAddEndPuncttrue
\mciteSetBstMidEndSepPunct{\mcitedefaultmidpunct}
{\mcitedefaultendpunct}{\mcitedefaultseppunct}\relax
\EndOfBibitem
\bibitem[Kardar(2007)]{kardar2007statistical}
Kardar,~M. \emph{Statistical physics of fields}; Cambridge University Press,
  2007\relax
\mciteBstWouldAddEndPuncttrue
\mciteSetBstMidEndSepPunct{\mcitedefaultmidpunct}
{\mcitedefaultendpunct}{\mcitedefaultseppunct}\relax
\EndOfBibitem
\bibitem[Fredrickson \latin{et~al.}(2002)Fredrickson, Ganesan, and
  Drolet]{fredrickson2002field}
Fredrickson,~G.~H.; Ganesan,~V.; Drolet,~F. Field-theoretic computer simulation
  methods for polymers and complex fluids. \emph{Macromolecules} \textbf{2002},
  \emph{35}, 16--39\relax
\mciteBstWouldAddEndPuncttrue
\mciteSetBstMidEndSepPunct{\mcitedefaultmidpunct}
{\mcitedefaultendpunct}{\mcitedefaultseppunct}\relax
\EndOfBibitem
\bibitem[Fredrickson(2006)]{fredrickson2006equilibrium}
Fredrickson,~G. \emph{The equilibrium theory of inhomogeneous polymers}; Oxford
  University Press, 2006\relax
\mciteBstWouldAddEndPuncttrue
\mciteSetBstMidEndSepPunct{\mcitedefaultmidpunct}
{\mcitedefaultendpunct}{\mcitedefaultseppunct}\relax
\EndOfBibitem
\bibitem[Fredrickson and Delaney(2023)Fredrickson, and
  Delaney]{fredrickson2023field}
Fredrickson,~G.~H.; Delaney,~K.~T. \emph{Field-Theoretic Simulations in Soft
  Matter and Quantum Fluids}; Oxford University Press, 2023; Vol. 173\relax
\mciteBstWouldAddEndPuncttrue
\mciteSetBstMidEndSepPunct{\mcitedefaultmidpunct}
{\mcitedefaultendpunct}{\mcitedefaultseppunct}\relax
\EndOfBibitem
\bibitem[Bernardi \latin{et~al.}(2015)Bernardi, Melo, and
  Schulten]{bernardi2015enhanced}
Bernardi,~R.~C.; Melo,~M.~C.; Schulten,~K. Enhanced sampling techniques in
  molecular dynamics simulations of biological systems. \emph{Biochimica et
  Biophysica Acta (BBA)-General Subjects} \textbf{2015}, \emph{1850},
  872--877\relax
\mciteBstWouldAddEndPuncttrue
\mciteSetBstMidEndSepPunct{\mcitedefaultmidpunct}
{\mcitedefaultendpunct}{\mcitedefaultseppunct}\relax
\EndOfBibitem
\bibitem[Valsson \latin{et~al.}(2016)Valsson, Tiwary, and
  Parrinello]{valsson2016enhancing}
Valsson,~O.; Tiwary,~P.; Parrinello,~M. Enhancing important fluctuations: Rare
  events and metadynamics from a conceptual viewpoint. \emph{Annual review of
  physical chemistry} \textbf{2016}, \emph{67}, 159--184\relax
\mciteBstWouldAddEndPuncttrue
\mciteSetBstMidEndSepPunct{\mcitedefaultmidpunct}
{\mcitedefaultendpunct}{\mcitedefaultseppunct}\relax
\EndOfBibitem
\bibitem[Sides \latin{et~al.}(2006)Sides, Kim, Kramer, and
  Fredrickson]{sides2006hybrid}
Sides,~S.~W.; Kim,~B.~J.; Kramer,~E.~J.; Fredrickson,~G.~H. Hybrid
  particle-field simulations of polymer nanocomposites. \emph{Physical review
  letters} \textbf{2006}, \emph{96}, 250601\relax
\mciteBstWouldAddEndPuncttrue
\mciteSetBstMidEndSepPunct{\mcitedefaultmidpunct}
{\mcitedefaultendpunct}{\mcitedefaultseppunct}\relax
\EndOfBibitem
\bibitem[Daoulas and Müller(2006)Daoulas, and Müller]{daoulas2006single}
Daoulas,~K.~C.; Müller,~M. {Single chain in mean field simulations:
  Quasi-instantaneous field approximation and quantitative comparison with
  Monte Carlo simulations}. \emph{The Journal of Chemical Physics}
  \textbf{2006}, \emph{125}, 184904\relax
\mciteBstWouldAddEndPuncttrue
\mciteSetBstMidEndSepPunct{\mcitedefaultmidpunct}
{\mcitedefaultendpunct}{\mcitedefaultseppunct}\relax
\EndOfBibitem
\bibitem[Milano and Kawakatsu(2009)Milano, and Kawakatsu]{milano2009hybrid}
Milano,~G.; Kawakatsu,~T. {Hybrid particle-field molecular dynamics simulations
  for dense polymer systems}. \emph{The Journal of Chemical Physics}
  \textbf{2009}, \emph{130}, 214106\relax
\mciteBstWouldAddEndPuncttrue
\mciteSetBstMidEndSepPunct{\mcitedefaultmidpunct}
{\mcitedefaultendpunct}{\mcitedefaultseppunct}\relax
\EndOfBibitem
\bibitem[M{\"u}ller(2011)]{muller2011studying}
M{\"u}ller,~M. Studying amphiphilic self-assembly with soft coarse-grained
  models. \emph{Journal of Statistical Physics} \textbf{2011}, \emph{145},
  967--1016\relax
\mciteBstWouldAddEndPuncttrue
\mciteSetBstMidEndSepPunct{\mcitedefaultmidpunct}
{\mcitedefaultendpunct}{\mcitedefaultseppunct}\relax
\EndOfBibitem
\bibitem[Edwards(1965)]{edwards1965statistical}
Edwards,~S.~F. The statistical mechanics of polymers with excluded volume.
  \emph{Proceedings of the Physical Society} \textbf{1965}, \emph{85},
  613\relax
\mciteBstWouldAddEndPuncttrue
\mciteSetBstMidEndSepPunct{\mcitedefaultmidpunct}
{\mcitedefaultendpunct}{\mcitedefaultseppunct}\relax
\EndOfBibitem
\bibitem[Edwards(1966)]{edwards1966theory}
Edwards,~S.~F. The theory of polymer solutions at intermediate concentration.
  \emph{Proceedings of the Physical Society} \textbf{1966}, \emph{88},
  265\relax
\mciteBstWouldAddEndPuncttrue
\mciteSetBstMidEndSepPunct{\mcitedefaultmidpunct}
{\mcitedefaultendpunct}{\mcitedefaultseppunct}\relax
\EndOfBibitem
\bibitem[Zhang \latin{et~al.}(2021)Zhang, Vigil, Sun, Bates, Loman, Murphy,
  Barbon, Song, Yu, Fredrickson, \latin{et~al.} others]{zhang2021emergence}
Zhang,~C.; Vigil,~D.~L.; Sun,~D.; Bates,~M.~W.; Loman,~T.; Murphy,~E.~A.;
  Barbon,~S.~M.; Song,~J.-A.; Yu,~B.; Fredrickson,~G.~H., \latin{et~al.}
  Emergence of hexagonally close-packed spheres in linear block copolymer
  melts. \emph{Journal of the American Chemical Society} \textbf{2021},
  \emph{143}, 14106--14114\relax
\mciteBstWouldAddEndPuncttrue
\mciteSetBstMidEndSepPunct{\mcitedefaultmidpunct}
{\mcitedefaultendpunct}{\mcitedefaultseppunct}\relax
\EndOfBibitem
\bibitem[Bates \latin{et~al.}(2019)Bates, Lequieu, Barbon, Lewis~III, Delaney,
  Anastasaki, Hawker, Fredrickson, and Bates]{bates2019stability}
Bates,~M.~W.; Lequieu,~J.; Barbon,~S.~M.; Lewis~III,~R.~M.; Delaney,~K.~T.;
  Anastasaki,~A.; Hawker,~C.~J.; Fredrickson,~G.~H.; Bates,~C.~M. Stability of
  the A15 phase in diblock copolymer melts. \emph{Proceedings of the National
  Academy of Sciences} \textbf{2019}, \emph{116}, 13194--13199\relax
\mciteBstWouldAddEndPuncttrue
\mciteSetBstMidEndSepPunct{\mcitedefaultmidpunct}
{\mcitedefaultendpunct}{\mcitedefaultseppunct}\relax
\EndOfBibitem
\bibitem[Wang and Voth(2005)Wang, and Voth]{wang2005unique}
Wang,~Y.; Voth,~G.~A. Unique spatial heterogeneity in ionic liquids.
  \emph{Journal of the American Chemical Society} \textbf{2005}, \emph{127},
  12192--12193\relax
\mciteBstWouldAddEndPuncttrue
\mciteSetBstMidEndSepPunct{\mcitedefaultmidpunct}
{\mcitedefaultendpunct}{\mcitedefaultseppunct}\relax
\EndOfBibitem
\bibitem[Hansen and McDonald(2013)Hansen, and McDonald]{hansen2013theory}
Hansen,~J.-P.; McDonald,~I.~R. \emph{Theory of simple liquids: with
  applications to soft matter}; Academic Press, 2013\relax
\mciteBstWouldAddEndPuncttrue
\mciteSetBstMidEndSepPunct{\mcitedefaultmidpunct}
{\mcitedefaultendpunct}{\mcitedefaultseppunct}\relax
\EndOfBibitem
\bibitem[Ramakrishnan and Yussouff(1979)Ramakrishnan, and
  Yussouff]{ramakrishnan1979first}
Ramakrishnan,~T.; Yussouff,~M. First-principles order-parameter theory of
  freezing. \emph{Physical Review B} \textbf{1979}, \emph{19}, 2775\relax
\mciteBstWouldAddEndPuncttrue
\mciteSetBstMidEndSepPunct{\mcitedefaultmidpunct}
{\mcitedefaultendpunct}{\mcitedefaultseppunct}\relax
\EndOfBibitem
\bibitem[Oxtoby(2002)]{oxtoby2002density}
Oxtoby,~D.~W. Density functional methods in the statistical mechanics of
  materials. \emph{Annual Review of Materials Research} \textbf{2002},
  \emph{32}, 39--52\relax
\mciteBstWouldAddEndPuncttrue
\mciteSetBstMidEndSepPunct{\mcitedefaultmidpunct}
{\mcitedefaultendpunct}{\mcitedefaultseppunct}\relax
\EndOfBibitem
\bibitem[Marconi and Tarazona(1999)Marconi, and Tarazona]{marconi1999dynamic}
Marconi,~U. M.~B.; Tarazona,~P. Dynamic density functional theory of fluids.
  \emph{The Journal of chemical physics} \textbf{1999}, \emph{110},
  8032--8044\relax
\mciteBstWouldAddEndPuncttrue
\mciteSetBstMidEndSepPunct{\mcitedefaultmidpunct}
{\mcitedefaultendpunct}{\mcitedefaultseppunct}\relax
\EndOfBibitem
\bibitem[Marconi and Tarazona(2000)Marconi, and Tarazona]{marconi2000dynamic}
Marconi,~U. M.~B.; Tarazona,~P. Dynamic density functional theory of fluids.
  \emph{Journal of Physics: Condensed Matter} \textbf{2000}, \emph{12},
  A413\relax
\mciteBstWouldAddEndPuncttrue
\mciteSetBstMidEndSepPunct{\mcitedefaultmidpunct}
{\mcitedefaultendpunct}{\mcitedefaultseppunct}\relax
\EndOfBibitem
\bibitem[Archer and Evans(2004)Archer, and Evans]{archer2004dynamical}
Archer,~A.~J.; Evans,~R. Dynamical density functional theory and its
  application to spinodal decomposition. \emph{The Journal of chemical physics}
  \textbf{2004}, \emph{121}, 4246--4254\relax
\mciteBstWouldAddEndPuncttrue
\mciteSetBstMidEndSepPunct{\mcitedefaultmidpunct}
{\mcitedefaultendpunct}{\mcitedefaultseppunct}\relax
\EndOfBibitem
\bibitem[Elder \latin{et~al.}(2007)Elder, Provatas, Berry, Stefanovic, and
  Grant]{elder2007phase}
Elder,~K.~R.; Provatas,~N.; Berry,~J.; Stefanovic,~P.; Grant,~M. Phase-field
  crystal modeling and classical density functional theory of freezing.
  \emph{Physical Review B} \textbf{2007}, \emph{75}, 064107\relax
\mciteBstWouldAddEndPuncttrue
\mciteSetBstMidEndSepPunct{\mcitedefaultmidpunct}
{\mcitedefaultendpunct}{\mcitedefaultseppunct}\relax
\EndOfBibitem
\bibitem[van Teeffelen \latin{et~al.}(2009)van Teeffelen, Backofen, Voigt, and
  L{\"o}wen]{van2009derivation}
van Teeffelen,~S.; Backofen,~R.; Voigt,~A.; L{\"o}wen,~H. Derivation of the
  phase-field-crystal model for colloidal solidification. \emph{Physical Review
  E} \textbf{2009}, \emph{79}, 051404\relax
\mciteBstWouldAddEndPuncttrue
\mciteSetBstMidEndSepPunct{\mcitedefaultmidpunct}
{\mcitedefaultendpunct}{\mcitedefaultseppunct}\relax
\EndOfBibitem
\bibitem[Emmerich \latin{et~al.}(2012)Emmerich, L{\"o}wen, Wittkowski, Gruhn,
  T{\'o}th, Tegze, and Gr{\'a}n{\'a}sy]{emmerich2012phase}
Emmerich,~H.; L{\"o}wen,~H.; Wittkowski,~R.; Gruhn,~T.; T{\'o}th,~G.~I.;
  Tegze,~G.; Gr{\'a}n{\'a}sy,~L. Phase-field-crystal models for condensed
  matter dynamics on atomic length and diffusive time scales: an overview.
  \emph{Advances in Physics} \textbf{2012}, \emph{61}, 665--743\relax
\mciteBstWouldAddEndPuncttrue
\mciteSetBstMidEndSepPunct{\mcitedefaultmidpunct}
{\mcitedefaultendpunct}{\mcitedefaultseppunct}\relax
\EndOfBibitem
\bibitem[Hubbard(1959)]{hubbard1959calculation}
Hubbard,~J. Calculation of partition functions. \emph{Physical Review Letters}
  \textbf{1959}, \emph{3}, 77\relax
\mciteBstWouldAddEndPuncttrue
\mciteSetBstMidEndSepPunct{\mcitedefaultmidpunct}
{\mcitedefaultendpunct}{\mcitedefaultseppunct}\relax
\EndOfBibitem
\bibitem[Fredrickson and Delaney(2018)Fredrickson, and
  Delaney]{fredrickson2018field}
Fredrickson,~G.~H.; Delaney,~K.~T. Field-theoretic simulations: An emerging
  tool for probing soft material assembly. \emph{MRS Bulletin} \textbf{2018},
  \emph{43}, 371--378\relax
\mciteBstWouldAddEndPuncttrue
\mciteSetBstMidEndSepPunct{\mcitedefaultmidpunct}
{\mcitedefaultendpunct}{\mcitedefaultseppunct}\relax
\EndOfBibitem
\bibitem[Singer(1958)]{singer1958use}
Singer,~K. Use of Gaussian Functions for Intermolecular Potentials.
  \emph{Nature} \textbf{1958}, \emph{181}, 262--263\relax
\mciteBstWouldAddEndPuncttrue
\mciteSetBstMidEndSepPunct{\mcitedefaultmidpunct}
{\mcitedefaultendpunct}{\mcitedefaultseppunct}\relax
\EndOfBibitem
\bibitem[Laird \latin{et~al.}(1991)Laird, Budimir, and
  Skinner]{laird1991quantum}
Laird,~B.~B.; Budimir,~J.; Skinner,~J.~L. Quantum-mechanical derivation of the
  Bloch equations: Beyond the weak-coupling limit. \emph{The Journal of
  chemical physics} \textbf{1991}, \emph{94}, 4391--4404\relax
\mciteBstWouldAddEndPuncttrue
\mciteSetBstMidEndSepPunct{\mcitedefaultmidpunct}
{\mcitedefaultendpunct}{\mcitedefaultseppunct}\relax
\EndOfBibitem
\bibitem[Bart{\'o}k \latin{et~al.}(2010)Bart{\'o}k, Payne, Kondor, and
  Cs{\'a}nyi]{bartok2010gaussian}
Bart{\'o}k,~A.~P.; Payne,~M.~C.; Kondor,~R.; Cs{\'a}nyi,~G. Gaussian
  approximation potentials: The accuracy of quantum mechanics, without the
  electrons. \emph{Physical review letters} \textbf{2010}, \emph{104},
  136403\relax
\mciteBstWouldAddEndPuncttrue
\mciteSetBstMidEndSepPunct{\mcitedefaultmidpunct}
{\mcitedefaultendpunct}{\mcitedefaultseppunct}\relax
\EndOfBibitem
\bibitem[John and Cs{\'a}nyi(2017)John, and Cs{\'a}nyi]{john2017many}
John,~S.; Cs{\'a}nyi,~G. Many-body coarse-grained interactions using Gaussian
  approximation potentials. \emph{The Journal of Physical Chemistry B}
  \textbf{2017}, \emph{121}, 10934--10949\relax
\mciteBstWouldAddEndPuncttrue
\mciteSetBstMidEndSepPunct{\mcitedefaultmidpunct}
{\mcitedefaultendpunct}{\mcitedefaultseppunct}\relax
\EndOfBibitem
\bibitem[Deringer \latin{et~al.}(2021)Deringer, Bart{\'o}k, Bernstein, Wilkins,
  Ceriotti, and Cs{\'a}nyi]{deringer2021gaussian}
Deringer,~V.~L.; Bart{\'o}k,~A.~P.; Bernstein,~N.; Wilkins,~D.~M.;
  Ceriotti,~M.; Cs{\'a}nyi,~G. Gaussian process regression for materials and
  molecules. \emph{Chemical Reviews} \textbf{2021}, \emph{121},
  10073--10141\relax
\mciteBstWouldAddEndPuncttrue
\mciteSetBstMidEndSepPunct{\mcitedefaultmidpunct}
{\mcitedefaultendpunct}{\mcitedefaultseppunct}\relax
\EndOfBibitem
\bibitem[Zwanzig(1954)]{zwanzig1954high}
Zwanzig,~R.~W. High-temperature equation of state by a perturbation method. I.
  Nonpolar gases. \emph{The Journal of Chemical Physics} \textbf{1954},
  \emph{22}, 1420--1426\relax
\mciteBstWouldAddEndPuncttrue
\mciteSetBstMidEndSepPunct{\mcitedefaultmidpunct}
{\mcitedefaultendpunct}{\mcitedefaultseppunct}\relax
\EndOfBibitem
\bibitem[Barker and Henderson(1967)Barker, and
  Henderson]{barker1967perturbation}
Barker,~J.~A.; Henderson,~D. Perturbation theory and equation of state for
  fluids. II. A successful theory of liquids. \emph{The Journal of chemical
  physics} \textbf{1967}, \emph{47}, 4714--4721\relax
\mciteBstWouldAddEndPuncttrue
\mciteSetBstMidEndSepPunct{\mcitedefaultmidpunct}
{\mcitedefaultendpunct}{\mcitedefaultseppunct}\relax
\EndOfBibitem
\bibitem[Weeks \latin{et~al.}(1971)Weeks, Chandler, and
  Andersen]{weeks1971role}
Weeks,~J.~D.; Chandler,~D.; Andersen,~H.~C. Role of repulsive forces in
  determining the equilibrium structure of simple liquids. \emph{The Journal of
  chemical physics} \textbf{1971}, \emph{54}, 5237--5247\relax
\mciteBstWouldAddEndPuncttrue
\mciteSetBstMidEndSepPunct{\mcitedefaultmidpunct}
{\mcitedefaultendpunct}{\mcitedefaultseppunct}\relax
\EndOfBibitem
\bibitem[Andersen and Chandler(1970)Andersen, and Chandler]{andersen1970mode}
Andersen,~H.~C.; Chandler,~D. Mode expansion in equilibrium statistical
  mechanics. I. General theory and application to the classical electron gas.
  \emph{The Journal of Chemical Physics} \textbf{1970}, \emph{53},
  547--554\relax
\mciteBstWouldAddEndPuncttrue
\mciteSetBstMidEndSepPunct{\mcitedefaultmidpunct}
{\mcitedefaultendpunct}{\mcitedefaultseppunct}\relax
\EndOfBibitem
\bibitem[Chandler and Andersen(1971)Chandler, and Andersen]{chandler1971mode}
Chandler,~D.; Andersen,~H.~C. Mode expansion in equilibrium statistical
  mechanics. II. A rapidly convergent theory of ionic solutions. \emph{The
  Journal of Chemical Physics} \textbf{1971}, \emph{54}, 26--33\relax
\mciteBstWouldAddEndPuncttrue
\mciteSetBstMidEndSepPunct{\mcitedefaultmidpunct}
{\mcitedefaultendpunct}{\mcitedefaultseppunct}\relax
\EndOfBibitem
\bibitem[Andersen and Chandler(1971)Andersen, and Chandler]{andersen1971mode}
Andersen,~H.~C.; Chandler,~D. Mode expansion in equilibrium statistical
  mechanics. III. Optimized convergence and application to ionic solution
  theory. \emph{The Journal of Chemical Physics} \textbf{1971}, \emph{55},
  1497--1504\relax
\mciteBstWouldAddEndPuncttrue
\mciteSetBstMidEndSepPunct{\mcitedefaultmidpunct}
{\mcitedefaultendpunct}{\mcitedefaultseppunct}\relax
\EndOfBibitem
\bibitem[Stillinger(1976)]{stillinger1976phase}
Stillinger,~F.~H. Phase transitions in the Gaussian core system. \emph{The
  Journal of Chemical Physics} \textbf{1976}, \emph{65}, 3968--3974\relax
\mciteBstWouldAddEndPuncttrue
\mciteSetBstMidEndSepPunct{\mcitedefaultmidpunct}
{\mcitedefaultendpunct}{\mcitedefaultseppunct}\relax
\EndOfBibitem
\bibitem[Yukawa(1935)]{yukawa1935interaction}
Yukawa,~H. On the interaction of elementary particles. I. \emph{Proceedings of
  the Physico-Mathematical Society of Japan. 3rd Series} \textbf{1935},
  \emph{17}, 48--57\relax
\mciteBstWouldAddEndPuncttrue
\mciteSetBstMidEndSepPunct{\mcitedefaultmidpunct}
{\mcitedefaultendpunct}{\mcitedefaultseppunct}\relax
\EndOfBibitem
\bibitem[Louis \latin{et~al.}(2000)Louis, Bolhuis, Hansen, and
  Meijer]{louis2000can}
Louis,~A.; Bolhuis,~P.; Hansen,~J.; Meijer,~E. Can polymer coils be modeled as
  “soft colloids”? \emph{Physical review letters} \textbf{2000}, \emph{85},
  2522\relax
\mciteBstWouldAddEndPuncttrue
\mciteSetBstMidEndSepPunct{\mcitedefaultmidpunct}
{\mcitedefaultendpunct}{\mcitedefaultseppunct}\relax
\EndOfBibitem
\bibitem[Ikeda and Miyazaki(2011)Ikeda, and Miyazaki]{ikeda2011glass}
Ikeda,~A.; Miyazaki,~K. Glass transition of the monodisperse Gaussian core
  model. \emph{Physical review letters} \textbf{2011}, \emph{106}, 015701\relax
\mciteBstWouldAddEndPuncttrue
\mciteSetBstMidEndSepPunct{\mcitedefaultmidpunct}
{\mcitedefaultendpunct}{\mcitedefaultseppunct}\relax
\EndOfBibitem
\bibitem[Ornstein and Zernike(1914)Ornstein, and
  Zernike]{ornstein1914influence}
Ornstein,~L.; Zernike,~F. The influence of accidental deviations of density on
  the equation of state. \emph{Koninklijke Nederlandsche Akademie van
  Wetenschappen Proceedings} \textbf{1914}, \emph{19}, 1312--1315\relax
\mciteBstWouldAddEndPuncttrue
\mciteSetBstMidEndSepPunct{\mcitedefaultmidpunct}
{\mcitedefaultendpunct}{\mcitedefaultseppunct}\relax
\EndOfBibitem
\bibitem[Morita(1958)]{morita1958theory}
Morita,~T. Theory of classical fluids: Hyper-netted chain approximation, I:
  Formulation for a one-component system. \emph{Progress of Theoretical
  Physics} \textbf{1958}, \emph{20}, 920--938\relax
\mciteBstWouldAddEndPuncttrue
\mciteSetBstMidEndSepPunct{\mcitedefaultmidpunct}
{\mcitedefaultendpunct}{\mcitedefaultseppunct}\relax
\EndOfBibitem
\bibitem[Percus and Yevick(1958)Percus, and Yevick]{percus1958analysis}
Percus,~J.~K.; Yevick,~G.~J. Analysis of classical statistical mechanics by
  means of collective coordinates. \emph{Physical Review} \textbf{1958},
  \emph{110}, 1\relax
\mciteBstWouldAddEndPuncttrue
\mciteSetBstMidEndSepPunct{\mcitedefaultmidpunct}
{\mcitedefaultendpunct}{\mcitedefaultseppunct}\relax
\EndOfBibitem
\bibitem[Sullivan and Gray(1981)Sullivan, and Gray]{sullivan1981evaluation}
Sullivan,~D.~E.; Gray,~C. Evaluation of angular correlation parameters and the
  dielectric constant in the RISM approximation. \emph{Molecular Physics}
  \textbf{1981}, \emph{42}, 443--454\relax
\mciteBstWouldAddEndPuncttrue
\mciteSetBstMidEndSepPunct{\mcitedefaultmidpunct}
{\mcitedefaultendpunct}{\mcitedefaultseppunct}\relax
\EndOfBibitem
\bibitem[Ho/ye and Stell(1976)Ho/ye, and Stell]{ho1976dielectric}
Ho/ye,~J.~S.; Stell,~G. Dielectric constant in terms of atom--atom correlation
  functions. \emph{The Journal of Chemical Physics} \textbf{1976}, \emph{65},
  18--22\relax
\mciteBstWouldAddEndPuncttrue
\mciteSetBstMidEndSepPunct{\mcitedefaultmidpunct}
{\mcitedefaultendpunct}{\mcitedefaultseppunct}\relax
\EndOfBibitem
\bibitem[Ho/ye and Stell(1977)Ho/ye, and Stell]{ho1977kerr}
Ho/ye,~J.; Stell,~G. Kerr constant and related quantities in terms of
  atom--atom correlation functions. \emph{The Journal of Chemical Physics}
  \textbf{1977}, \emph{66}, 795--801\relax
\mciteBstWouldAddEndPuncttrue
\mciteSetBstMidEndSepPunct{\mcitedefaultmidpunct}
{\mcitedefaultendpunct}{\mcitedefaultseppunct}\relax
\EndOfBibitem
\bibitem[Jin \latin{et~al.}(2019)Jin, Pak, and Voth]{jin2019understanding}
Jin,~J.; Pak,~A.~J.; Voth,~G.~A. Understanding missing entropy in
  coarse-grained systems: Addressing issues of representability and
  transferability. \emph{The journal of physical chemistry letters}
  \textbf{2019}, \emph{10}, 4549--4557\relax
\mciteBstWouldAddEndPuncttrue
\mciteSetBstMidEndSepPunct{\mcitedefaultmidpunct}
{\mcitedefaultendpunct}{\mcitedefaultseppunct}\relax
\EndOfBibitem
\bibitem[Dannenhoffer-Lafage \latin{et~al.}(2019)Dannenhoffer-Lafage, Wagner,
  Durumeric, and Voth]{dannenhoffer2019compatible}
Dannenhoffer-Lafage,~T.; Wagner,~J.~W.; Durumeric,~A. E.~P.; Voth,~G.~A.
  {Compatible observable decompositions for coarse-grained representations of
  real molecular systems}. \emph{The Journal of Chemical Physics}
  \textbf{2019}, \emph{151}, 134115\relax
\mciteBstWouldAddEndPuncttrue
\mciteSetBstMidEndSepPunct{\mcitedefaultmidpunct}
{\mcitedefaultendpunct}{\mcitedefaultseppunct}\relax
\EndOfBibitem
\bibitem[Johnson \latin{et~al.}(2007)Johnson, Head-Gordon, and
  Louis]{johnson2007representability}
Johnson,~M.~E.; Head-Gordon,~T.; Louis,~A.~A. {Representability problems for
  coarse-grained water potentials}. \emph{The Journal of Chemical Physics}
  \textbf{2007}, \emph{126}, 144509\relax
\mciteBstWouldAddEndPuncttrue
\mciteSetBstMidEndSepPunct{\mcitedefaultmidpunct}
{\mcitedefaultendpunct}{\mcitedefaultseppunct}\relax
\EndOfBibitem
\bibitem[Wagner \latin{et~al.}(2016)Wagner, Dama, Durumeric, and
  Voth]{wagner2016on}
Wagner,~J.~W.; Dama,~J.~F.; Durumeric,~A. E.~P.; Voth,~G.~A. {On the
  representability problem and the physical meaning of coarse-grained models}.
  \emph{The Journal of Chemical Physics} \textbf{2016}, \emph{145},
  044108\relax
\mciteBstWouldAddEndPuncttrue
\mciteSetBstMidEndSepPunct{\mcitedefaultmidpunct}
{\mcitedefaultendpunct}{\mcitedefaultseppunct}\relax
\EndOfBibitem
\bibitem[Dunn \latin{et~al.}(2016)Dunn, Foley, and Noid]{dunn2016van}
Dunn,~N.~J.; Foley,~T.~T.; Noid,~W.~G. Van der Waals perspective on
  coarse-graining: Progress toward solving representability and transferability
  problems. \emph{Accounts of Chemical Research} \textbf{2016}, \emph{49},
  2832--2840\relax
\mciteBstWouldAddEndPuncttrue
\mciteSetBstMidEndSepPunct{\mcitedefaultmidpunct}
{\mcitedefaultendpunct}{\mcitedefaultseppunct}\relax
\EndOfBibitem
\bibitem[Mart{\'\i}nez \latin{et~al.}(2009)Mart{\'\i}nez, Andrade, Birgin, and
  Mart{\'\i}nez]{martinez2009packmol}
Mart{\'\i}nez,~L.; Andrade,~R.; Birgin,~E.~G.; Mart{\'\i}nez,~J.~M. PACKMOL: A
  package for building initial configurations for molecular dynamics
  simulations. \emph{Journal of computational chemistry} \textbf{2009},
  \emph{30}, 2157--2164\relax
\mciteBstWouldAddEndPuncttrue
\mciteSetBstMidEndSepPunct{\mcitedefaultmidpunct}
{\mcitedefaultendpunct}{\mcitedefaultseppunct}\relax
\EndOfBibitem
\bibitem[Nos{\'e}(1984)]{nose1984unified}
Nos{\'e},~S. A unified formulation of the constant temperature molecular
  dynamics methods. \emph{The Journal of chemical physics} \textbf{1984},
  \emph{81}, 511--519\relax
\mciteBstWouldAddEndPuncttrue
\mciteSetBstMidEndSepPunct{\mcitedefaultmidpunct}
{\mcitedefaultendpunct}{\mcitedefaultseppunct}\relax
\EndOfBibitem
\bibitem[Hoover(1985)]{hoover1985canonical}
Hoover,~W.~G. Canonical dynamics: Equilibrium phase-space distributions.
  \emph{Physical review A} \textbf{1985}, \emph{31}, 1695\relax
\mciteBstWouldAddEndPuncttrue
\mciteSetBstMidEndSepPunct{\mcitedefaultmidpunct}
{\mcitedefaultendpunct}{\mcitedefaultseppunct}\relax
\EndOfBibitem
\bibitem[Plimpton(1995)]{plimpton1995fast}
Plimpton,~S. Fast parallel algorithms for short-range molecular dynamics.
  \emph{Journal of computational physics} \textbf{1995}, \emph{117},
  1--19\relax
\mciteBstWouldAddEndPuncttrue
\mciteSetBstMidEndSepPunct{\mcitedefaultmidpunct}
{\mcitedefaultendpunct}{\mcitedefaultseppunct}\relax
\EndOfBibitem
\bibitem[Brown \latin{et~al.}(2011)Brown, Wang, Plimpton, and
  Tharrington]{brown2011implementing}
Brown,~W.~M.; Wang,~P.; Plimpton,~S.~J.; Tharrington,~A.~N. Implementing
  molecular dynamics on hybrid high performance computers--short range forces.
  \emph{Computer Physics Communications} \textbf{2011}, \emph{182},
  898--911\relax
\mciteBstWouldAddEndPuncttrue
\mciteSetBstMidEndSepPunct{\mcitedefaultmidpunct}
{\mcitedefaultendpunct}{\mcitedefaultseppunct}\relax
\EndOfBibitem
\bibitem[Brown \latin{et~al.}(2012)Brown, Kohlmeyer, Plimpton, and
  Tharrington]{brown2012implementing}
Brown,~W.~M.; Kohlmeyer,~A.; Plimpton,~S.~J.; Tharrington,~A.~N. Implementing
  molecular dynamics on hybrid high performance computers--Particle--particle
  particle-mesh. \emph{Computer Physics Communications} \textbf{2012},
  \emph{183}, 449--459\relax
\mciteBstWouldAddEndPuncttrue
\mciteSetBstMidEndSepPunct{\mcitedefaultmidpunct}
{\mcitedefaultendpunct}{\mcitedefaultseppunct}\relax
\EndOfBibitem
\bibitem[Levesque(1966)]{levesque1966etude}
Levesque,~D. Etude des equations de percus et yevick, d'hyperchaine et de born
  et green dans le cas des fluides classiques. \emph{Physica} \textbf{1966},
  \emph{32}, 1985--2006\relax
\mciteBstWouldAddEndPuncttrue
\mciteSetBstMidEndSepPunct{\mcitedefaultmidpunct}
{\mcitedefaultendpunct}{\mcitedefaultseppunct}\relax
\EndOfBibitem
\bibitem[Allinger \latin{et~al.}(1989)Allinger, Yuh, and
  Lii]{allinger1989molecular}
Allinger,~N.~L.; Yuh,~Y.~H.; Lii,~J.~H. Molecular mechanics. The MM3 force
  field for hydrocarbons. 1. \emph{Journal of the American Chemical Society}
  \textbf{1989}, \emph{111}, 8551--8566\relax
\mciteBstWouldAddEndPuncttrue
\mciteSetBstMidEndSepPunct{\mcitedefaultmidpunct}
{\mcitedefaultendpunct}{\mcitedefaultseppunct}\relax
\EndOfBibitem
\bibitem[Mayo \latin{et~al.}(1990)Mayo, Olafson, and Goddard]{mayo1990dreiding}
Mayo,~S.~L.; Olafson,~B.~D.; Goddard,~W.~A. DREIDING: a generic force field for
  molecular simulations. \emph{Journal of Physical chemistry} \textbf{1990},
  \emph{94}, 8897--8909\relax
\mciteBstWouldAddEndPuncttrue
\mciteSetBstMidEndSepPunct{\mcitedefaultmidpunct}
{\mcitedefaultendpunct}{\mcitedefaultseppunct}\relax
\EndOfBibitem
\bibitem[Rapp{\'e} \latin{et~al.}(1992)Rapp{\'e}, Casewit, Colwell,
  Goddard~III, and Skiff]{rappe1992uff}
Rapp{\'e},~A.~K.; Casewit,~C.~J.; Colwell,~K.; Goddard~III,~W.~A.; Skiff,~W.~M.
  UFF, a full periodic table force field for molecular mechanics and molecular
  dynamics simulations. \emph{Journal of the American chemical society}
  \textbf{1992}, \emph{114}, 10024--10035\relax
\mciteBstWouldAddEndPuncttrue
\mciteSetBstMidEndSepPunct{\mcitedefaultmidpunct}
{\mcitedefaultendpunct}{\mcitedefaultseppunct}\relax
\EndOfBibitem
\bibitem[Scott \latin{et~al.}(1999)Scott, H{\"u}nenberger, Tironi, Mark,
  Billeter, Fennen, Torda, Huber, Kr{\"u}ger, and
  Van~Gunsteren]{scott1999gromos}
Scott,~W.~R.; H{\"u}nenberger,~P.~H.; Tironi,~I.~G.; Mark,~A.~E.;
  Billeter,~S.~R.; Fennen,~J.; Torda,~A.~E.; Huber,~T.; Kr{\"u}ger,~P.;
  Van~Gunsteren,~W.~F. The GROMOS biomolecular simulation program package.
  \emph{The Journal of Physical Chemistry A} \textbf{1999}, \emph{103},
  3596--3607\relax
\mciteBstWouldAddEndPuncttrue
\mciteSetBstMidEndSepPunct{\mcitedefaultmidpunct}
{\mcitedefaultendpunct}{\mcitedefaultseppunct}\relax
\EndOfBibitem
\bibitem[Case \latin{et~al.}(2005)Case, Cheatham~III, Darden, Gohlke, Luo,
  Merz~Jr, Onufriev, Simmerling, Wang, and Woods]{case2005amber}
Case,~D.~A.; Cheatham~III,~T.~E.; Darden,~T.; Gohlke,~H.; Luo,~R.;
  Merz~Jr,~K.~M.; Onufriev,~A.; Simmerling,~C.; Wang,~B.; Woods,~R.~J. The
  Amber biomolecular simulation programs. \emph{Journal of computational
  chemistry} \textbf{2005}, \emph{26}, 1668--1688\relax
\mciteBstWouldAddEndPuncttrue
\mciteSetBstMidEndSepPunct{\mcitedefaultmidpunct}
{\mcitedefaultendpunct}{\mcitedefaultseppunct}\relax
\EndOfBibitem
\bibitem[Brooks \latin{et~al.}(2009)Brooks, Brooks~III, Mackerell~Jr, Nilsson,
  Petrella, Roux, Won, Archontis, Bartels, Boresch, \latin{et~al.}
  others]{brooks2009charmm}
Brooks,~B.~R.; Brooks~III,~C.~L.; Mackerell~Jr,~A.~D.; Nilsson,~L.;
  Petrella,~R.~J.; Roux,~B.; Won,~Y.; Archontis,~G.; Bartels,~C.; Boresch,~S.,
  \latin{et~al.}  CHARMM: the biomolecular simulation program. \emph{Journal of
  computational chemistry} \textbf{2009}, \emph{30}, 1545--1614\relax
\mciteBstWouldAddEndPuncttrue
\mciteSetBstMidEndSepPunct{\mcitedefaultmidpunct}
{\mcitedefaultendpunct}{\mcitedefaultseppunct}\relax
\EndOfBibitem
\bibitem[Noid \latin{et~al.}(2008)Noid, Chu, Ayton, Krishna, Izvekov, Voth,
  Das, and Andersen]{noid2008multiscale1}
Noid,~W.~G.; Chu,~J.-W.; Ayton,~G.~S.; Krishna,~V.; Izvekov,~S.; Voth,~G.~A.;
  Das,~A.; Andersen,~H.~C. {The multiscale coarse-graining method. I. A
  rigorous bridge between atomistic and coarse-grained models}. \emph{The
  Journal of Chemical Physics} \textbf{2008}, \emph{128}, 244114\relax
\mciteBstWouldAddEndPuncttrue
\mciteSetBstMidEndSepPunct{\mcitedefaultmidpunct}
{\mcitedefaultendpunct}{\mcitedefaultseppunct}\relax
\EndOfBibitem
\bibitem[Noid \latin{et~al.}(2008)Noid, Liu, Wang, Chu, Ayton, Izvekov,
  Andersen, and Voth]{noid2008multiscale2}
Noid,~W.~G.; Liu,~P.; Wang,~Y.; Chu,~J.-W.; Ayton,~G.~S.; Izvekov,~S.;
  Andersen,~H.~C.; Voth,~G.~A. {The multiscale coarse-graining method. II.
  Numerical implementation for coarse-grained molecular models}. \emph{The
  Journal of Chemical Physics} \textbf{2008}, \emph{128}, 244115\relax
\mciteBstWouldAddEndPuncttrue
\mciteSetBstMidEndSepPunct{\mcitedefaultmidpunct}
{\mcitedefaultendpunct}{\mcitedefaultseppunct}\relax
\EndOfBibitem
\bibitem[Jin and Voth(2018)Jin, and Voth]{jin2018ultra}
Jin,~J.; Voth,~G.~A. Ultra-coarse-grained models allow for an accurate and
  transferable treatment of interfacial systems. \emph{Journal of chemical
  theory and computation} \textbf{2018}, \emph{14}, 2180--2197\relax
\mciteBstWouldAddEndPuncttrue
\mciteSetBstMidEndSepPunct{\mcitedefaultmidpunct}
{\mcitedefaultendpunct}{\mcitedefaultseppunct}\relax
\EndOfBibitem
\bibitem[Jin \latin{et~al.}(2020)Jin, Yu, and Voth]{jin2020temperature}
Jin,~J.; Yu,~A.; Voth,~G.~A. Temperature and phase transferable bottom-up
  coarse-grained models. \emph{Journal of chemical theory and computation}
  \textbf{2020}, \emph{16}, 6823--6842\relax
\mciteBstWouldAddEndPuncttrue
\mciteSetBstMidEndSepPunct{\mcitedefaultmidpunct}
{\mcitedefaultendpunct}{\mcitedefaultseppunct}\relax
\EndOfBibitem
\bibitem[Jorgensen \latin{et~al.}(1996)Jorgensen, Maxwell, and
  Tirado-Rives]{jorgensen1996development}
Jorgensen,~W.~L.; Maxwell,~D.~S.; Tirado-Rives,~J. Development and testing of
  the OPLS all-atom force field on conformational energetics and properties of
  organic liquids. \emph{Journal of the American Chemical Society}
  \textbf{1996}, \emph{118}, 11225--11236\relax
\mciteBstWouldAddEndPuncttrue
\mciteSetBstMidEndSepPunct{\mcitedefaultmidpunct}
{\mcitedefaultendpunct}{\mcitedefaultseppunct}\relax
\EndOfBibitem
\bibitem[Kaminski \latin{et~al.}(2001)Kaminski, Friesner, Tirado-Rives, and
  Jorgensen]{kaminski2001evaluation}
Kaminski,~G.~A.; Friesner,~R.~A.; Tirado-Rives,~J.; Jorgensen,~W.~L. Evaluation
  and reparametrization of the OPLS-AA force field for proteins via comparison
  with accurate quantum chemical calculations on peptides. \emph{The Journal of
  Physical Chemistry B} \textbf{2001}, \emph{105}, 6474--6487\relax
\mciteBstWouldAddEndPuncttrue
\mciteSetBstMidEndSepPunct{\mcitedefaultmidpunct}
{\mcitedefaultendpunct}{\mcitedefaultseppunct}\relax
\EndOfBibitem
\bibitem[Jin \latin{et~al.}(2021)Jin, Han, Pak, and Voth]{jin2021new1}
Jin,~J.; Han,~Y.; Pak,~A.~J.; Voth,~G.~A. {A new one-site coarse-grained model
  for water: Bottom-up many-body projected water (BUMPer). I. General theory
  and model}. \emph{The Journal of Chemical Physics} \textbf{2021}, \emph{154},
  044104\relax
\mciteBstWouldAddEndPuncttrue
\mciteSetBstMidEndSepPunct{\mcitedefaultmidpunct}
{\mcitedefaultendpunct}{\mcitedefaultseppunct}\relax
\EndOfBibitem
\bibitem[Jin \latin{et~al.}(2021)Jin, Pak, Han, and Voth]{jin2021new2}
Jin,~J.; Pak,~A.~J.; Han,~Y.; Voth,~G.~A. {A new one-site coarse-grained model
  for water: Bottom-up many-body projected water (BUMPer). II. Temperature
  transferability and structural properties at low temperature}. \emph{The
  Journal of Chemical Physics} \textbf{2021}, \emph{154}, 044105\relax
\mciteBstWouldAddEndPuncttrue
\mciteSetBstMidEndSepPunct{\mcitedefaultmidpunct}
{\mcitedefaultendpunct}{\mcitedefaultseppunct}\relax
\EndOfBibitem
\bibitem[Wu \latin{et~al.}(2006)Wu, Tepper, and Voth]{wu2006flexible}
Wu,~Y.; Tepper,~H.~L.; Voth,~G.~A. {Flexible simple point-charge water model
  with improved liquid-state properties}. \emph{The Journal of Chemical
  Physics} \textbf{2006}, \emph{124}, 024503\relax
\mciteBstWouldAddEndPuncttrue
\mciteSetBstMidEndSepPunct{\mcitedefaultmidpunct}
{\mcitedefaultendpunct}{\mcitedefaultseppunct}\relax
\EndOfBibitem
\bibitem[Lu \latin{et~al.}(2010)Lu, Izvekov, Das, Andersen, and
  Voth]{lu2010efficient}
Lu,~L.; Izvekov,~S.; Das,~A.; Andersen,~H.~C.; Voth,~G.~A. Efficient,
  regularized, and scalable algorithms for multiscale coarse-graining.
  \emph{Journal of chemical theory and computation} \textbf{2010}, \emph{6},
  954--965\relax
\mciteBstWouldAddEndPuncttrue
\mciteSetBstMidEndSepPunct{\mcitedefaultmidpunct}
{\mcitedefaultendpunct}{\mcitedefaultseppunct}\relax
\EndOfBibitem
\bibitem[Golubkov and Ren(2006)Golubkov, and Ren]{golubkov2006generalized}
Golubkov,~P.~A.; Ren,~P. {Generalized coarse-grained model based on point
  multipole and Gay-Berne potentials}. \emph{The Journal of Chemical Physics}
  \textbf{2006}, \emph{125}, 064103\relax
\mciteBstWouldAddEndPuncttrue
\mciteSetBstMidEndSepPunct{\mcitedefaultmidpunct}
{\mcitedefaultendpunct}{\mcitedefaultseppunct}\relax
\EndOfBibitem
\bibitem[Tait and Franks(1971)Tait, and Franks]{tait1971water}
Tait,~M.; Franks,~F. Water in biological systems. \emph{Nature} \textbf{1971},
  \emph{230}, 91--94\relax
\mciteBstWouldAddEndPuncttrue
\mciteSetBstMidEndSepPunct{\mcitedefaultmidpunct}
{\mcitedefaultendpunct}{\mcitedefaultseppunct}\relax
\EndOfBibitem
\bibitem[Israelachvili and Wennerstr{\"o}m(1996)Israelachvili, and
  Wennerstr{\"o}m]{israelachvili1996role}
Israelachvili,~J.; Wennerstr{\"o}m,~H. Role of hydration and water structure in
  biological and colloidal interactions. \emph{Nature} \textbf{1996},
  \emph{379}, 219--225\relax
\mciteBstWouldAddEndPuncttrue
\mciteSetBstMidEndSepPunct{\mcitedefaultmidpunct}
{\mcitedefaultendpunct}{\mcitedefaultseppunct}\relax
\EndOfBibitem
\bibitem[Dill \latin{et~al.}(2005)Dill, Truskett, Vlachy, and
  Hribar-Lee]{dill2005modeling}
Dill,~K.~A.; Truskett,~T.~M.; Vlachy,~V.; Hribar-Lee,~B. Modeling water, the
  hydrophobic effect, and ion solvation. \emph{Annu. Rev. Biophys. Biomol.
  Struct.} \textbf{2005}, \emph{34}, 173--199\relax
\mciteBstWouldAddEndPuncttrue
\mciteSetBstMidEndSepPunct{\mcitedefaultmidpunct}
{\mcitedefaultendpunct}{\mcitedefaultseppunct}\relax
\EndOfBibitem
\bibitem[Davidson and Franks(1973)Davidson, and Franks]{davidson1973water}
Davidson,~D.; Franks,~F. \emph{Water: a comprehensive treatise}; Plenum Press
  New York, 1973; Vol.~2; pp 115--234\relax
\mciteBstWouldAddEndPuncttrue
\mciteSetBstMidEndSepPunct{\mcitedefaultmidpunct}
{\mcitedefaultendpunct}{\mcitedefaultseppunct}\relax
\EndOfBibitem
\bibitem[Eisenberg and Kauzmann(2005)Eisenberg, and
  Kauzmann]{eisenberg2005structure}
Eisenberg,~D.; Kauzmann,~W. \emph{The structure and properties of water}; OUP
  Oxford, 2005\relax
\mciteBstWouldAddEndPuncttrue
\mciteSetBstMidEndSepPunct{\mcitedefaultmidpunct}
{\mcitedefaultendpunct}{\mcitedefaultseppunct}\relax
\EndOfBibitem
\bibitem[Hankins \latin{et~al.}(1970)Hankins, Moskowitz, and
  Stillinger]{hankins1970water}
Hankins,~D.; Moskowitz,~J.; Stillinger,~F. Water molecule interactions.
  \emph{The Journal of Chemical Physics} \textbf{1970}, \emph{53},
  4544--4554\relax
\mciteBstWouldAddEndPuncttrue
\mciteSetBstMidEndSepPunct{\mcitedefaultmidpunct}
{\mcitedefaultendpunct}{\mcitedefaultseppunct}\relax
\EndOfBibitem
\bibitem[Morita and Hiroike(1960)Morita, and Hiroike]{morita1960new}
Morita,~T.; Hiroike,~K. A new approach to the theory of classical fluids. I.
  \emph{Progress of Theoretical Physics} \textbf{1960}, \emph{23},
  1003--1027\relax
\mciteBstWouldAddEndPuncttrue
\mciteSetBstMidEndSepPunct{\mcitedefaultmidpunct}
{\mcitedefaultendpunct}{\mcitedefaultseppunct}\relax
\EndOfBibitem
\bibitem[Duh and Henderson(1996)Duh, and Henderson]{duh1996integral}
Duh,~D.-M.; Henderson,~D. Integral equation theory for Lennard-Jones fluids:
  The bridge function and applications to pure fluids and mixtures. \emph{The
  Journal of chemical physics} \textbf{1996}, \emph{104}, 6742--6754\relax
\mciteBstWouldAddEndPuncttrue
\mciteSetBstMidEndSepPunct{\mcitedefaultmidpunct}
{\mcitedefaultendpunct}{\mcitedefaultseppunct}\relax
\EndOfBibitem
\bibitem[Izvekov and Voth(2005)Izvekov, and Voth]{izvekov2005multiscale1}
Izvekov,~S.; Voth,~G.~A. {Multiscale coarse graining of liquid-state systems}.
  \emph{The Journal of Chemical Physics} \textbf{2005}, \emph{123},
  134105\relax
\mciteBstWouldAddEndPuncttrue
\mciteSetBstMidEndSepPunct{\mcitedefaultmidpunct}
{\mcitedefaultendpunct}{\mcitedefaultseppunct}\relax
\EndOfBibitem
\bibitem[Izvekov and Voth(2005)Izvekov, and Voth]{izvekov2005multiscale2}
Izvekov,~S.; Voth,~G.~A. A multiscale coarse-graining method for biomolecular
  systems. \emph{The Journal of Physical Chemistry B} \textbf{2005},
  \emph{109}, 2469--2473\relax
\mciteBstWouldAddEndPuncttrue
\mciteSetBstMidEndSepPunct{\mcitedefaultmidpunct}
{\mcitedefaultendpunct}{\mcitedefaultseppunct}\relax
\EndOfBibitem
\bibitem[Reith \latin{et~al.}(2003)Reith, P{\"u}tz, and
  M{\"u}ller-Plathe]{reith2003deriving}
Reith,~D.; P{\"u}tz,~M.; M{\"u}ller-Plathe,~F. Deriving effective mesoscale
  potentials from atomistic simulations. \emph{Journal of computational
  chemistry} \textbf{2003}, \emph{24}, 1624--1636\relax
\mciteBstWouldAddEndPuncttrue
\mciteSetBstMidEndSepPunct{\mcitedefaultmidpunct}
{\mcitedefaultendpunct}{\mcitedefaultseppunct}\relax
\EndOfBibitem
\bibitem[Shell(2008)]{shell2008relative}
Shell,~M.~S. {The relative entropy is fundamental to multiscale and inverse
  thermodynamic problems}. \emph{The Journal of Chemical Physics}
  \textbf{2008}, \emph{129}, 144108\relax
\mciteBstWouldAddEndPuncttrue
\mciteSetBstMidEndSepPunct{\mcitedefaultmidpunct}
{\mcitedefaultendpunct}{\mcitedefaultseppunct}\relax
\EndOfBibitem
\bibitem[Chaimovich and Shell(2010)Chaimovich, and
  Shell]{chaimovich2010relative}
Chaimovich,~A.; Shell,~M.~S. Relative entropy as a universal metric for
  multiscale errors. \emph{Physical Review E} \textbf{2010}, \emph{81},
  060104\relax
\mciteBstWouldAddEndPuncttrue
\mciteSetBstMidEndSepPunct{\mcitedefaultmidpunct}
{\mcitedefaultendpunct}{\mcitedefaultseppunct}\relax
\EndOfBibitem
\bibitem[Chaimovich and Shell(2011)Chaimovich, and Shell]{chaimovich2011coarse}
Chaimovich,~A.; Shell,~M.~S. {Coarse-graining errors and numerical optimization
  using a relative entropy framework}. \emph{The Journal of Chemical Physics}
  \textbf{2011}, \emph{134}, 094112\relax
\mciteBstWouldAddEndPuncttrue
\mciteSetBstMidEndSepPunct{\mcitedefaultmidpunct}
{\mcitedefaultendpunct}{\mcitedefaultseppunct}\relax
\EndOfBibitem
\bibitem[Shell(2016)]{shell2016coarse}
Shell,~M.~S. Coarse-Graining with the Relative Entropy. \emph{Advances in
  chemical physics} \textbf{2016}, \emph{161}, 395--441\relax
\mciteBstWouldAddEndPuncttrue
\mciteSetBstMidEndSepPunct{\mcitedefaultmidpunct}
{\mcitedefaultendpunct}{\mcitedefaultseppunct}\relax
\EndOfBibitem
\bibitem[Soper(1996)]{soper1996empirical}
Soper,~A. Empirical potential Monte Carlo simulation of fluid structure.
  \emph{Chemical Physics} \textbf{1996}, \emph{202}, 295--306\relax
\mciteBstWouldAddEndPuncttrue
\mciteSetBstMidEndSepPunct{\mcitedefaultmidpunct}
{\mcitedefaultendpunct}{\mcitedefaultseppunct}\relax
\EndOfBibitem
\bibitem[Lyubartsev and Laaksonen(1995)Lyubartsev, and
  Laaksonen]{lyubartsev1995calculation}
Lyubartsev,~A.~P.; Laaksonen,~A. Calculation of effective interaction
  potentials from radial distribution functions: A reverse Monte Carlo
  approach. \emph{Physical Review E} \textbf{1995}, \emph{52}, 3730\relax
\mciteBstWouldAddEndPuncttrue
\mciteSetBstMidEndSepPunct{\mcitedefaultmidpunct}
{\mcitedefaultendpunct}{\mcitedefaultseppunct}\relax
\EndOfBibitem
\bibitem[Kullback and Leibler(1951)Kullback, and
  Leibler]{kullback1951information}
Kullback,~S.; Leibler,~R.~A. On information and sufficiency. \emph{The annals
  of mathematical statistics} \textbf{1951}, \emph{22}, 79--86\relax
\mciteBstWouldAddEndPuncttrue
\mciteSetBstMidEndSepPunct{\mcitedefaultmidpunct}
{\mcitedefaultendpunct}{\mcitedefaultseppunct}\relax
\EndOfBibitem
\bibitem[Peng \latin{et~al.}(2023)Peng, Pak, Durumeric, Sahrmann, Mani, Jin,
  Loose, Beiter, and Voth]{OpenMSCG}
Peng,~Y.; Pak,~A.~J.; Durumeric,~A. E.~P.; Sahrmann,~P.~G.; Mani,~S.; Jin,~J.;
  Loose,~T.~D.; Beiter,~J.; Voth,~G.~A. OpenMSCG: A Software Tool for Bottom-Up
  Coarse-Graining. \emph{The Journal of Physical Chemistry B} \textbf{2023},
  \emph{127}, 8537--8550\relax
\mciteBstWouldAddEndPuncttrue
\mciteSetBstMidEndSepPunct{\mcitedefaultmidpunct}
{\mcitedefaultendpunct}{\mcitedefaultseppunct}\relax
\EndOfBibitem
\bibitem[Mayer and Mayer(1940)Mayer, and Mayer]{mayer1940statistical}
Mayer,~J.~E.; Mayer,~M.~G. \emph{Statistical mechanics}; John Wiley \& Sons New
  York, 1940; Vol.~28\relax
\mciteBstWouldAddEndPuncttrue
\mciteSetBstMidEndSepPunct{\mcitedefaultmidpunct}
{\mcitedefaultendpunct}{\mcitedefaultseppunct}\relax
\EndOfBibitem
\bibitem[Rosenfeld and Ashcroft(1979)Rosenfeld, and
  Ashcroft]{rosenfeld1979theory}
Rosenfeld,~Y.; Ashcroft,~N. Theory of simple classical fluids: Universality in
  the short-range structure. \emph{Physical Review A} \textbf{1979}, \emph{20},
  1208\relax
\mciteBstWouldAddEndPuncttrue
\mciteSetBstMidEndSepPunct{\mcitedefaultmidpunct}
{\mcitedefaultendpunct}{\mcitedefaultseppunct}\relax
\EndOfBibitem
\bibitem[Llano-Restrepo and Chapman(1992)Llano-Restrepo, and
  Chapman]{llano1992bridge}
Llano-Restrepo,~M.; Chapman,~W.~G. Bridge function and cavity correlation
  function for the Lennard-Jones fluid from simulation. \emph{The Journal of
  chemical physics} \textbf{1992}, \emph{97}, 2046--2054\relax
\mciteBstWouldAddEndPuncttrue
\mciteSetBstMidEndSepPunct{\mcitedefaultmidpunct}
{\mcitedefaultendpunct}{\mcitedefaultseppunct}\relax
\EndOfBibitem
\bibitem[Dean(1996)]{dean1996langevin}
Dean,~D.~S. Langevin equation for the density of a system of interacting
  Langevin processes. \emph{Journal of Physics A: Mathematical and General}
  \textbf{1996}, \emph{29}, L613\relax
\mciteBstWouldAddEndPuncttrue
\mciteSetBstMidEndSepPunct{\mcitedefaultmidpunct}
{\mcitedefaultendpunct}{\mcitedefaultseppunct}\relax
\EndOfBibitem
\bibitem[Kawasaki(1998)]{kawasaki1998microscopic}
Kawasaki,~K. Microscopic analyses of the dynamical density functional equation
  of dense fluids. \emph{Journal of statistical physics} \textbf{1998},
  \emph{93}, 527--546\relax
\mciteBstWouldAddEndPuncttrue
\mciteSetBstMidEndSepPunct{\mcitedefaultmidpunct}
{\mcitedefaultendpunct}{\mcitedefaultseppunct}\relax
\EndOfBibitem
\end{mcitethebibliography}

\end{document}